\documentclass{article}

\usepackage{subcaption}
\usepackage{graphicx}
\usepackage{amsmath}
\usepackage{amssymb}
\usepackage{mathrsfs}
\usepackage{amsthm}
\usepackage{mathtools}

\usepackage[main,final,nonatbib]{neurips_2026}
\usepackage[utf8]{inputenc} % allow utf-8 input
\usepackage[T1]{fontenc}    % use 8-bit T1 fonts
\usepackage[colorlinks=true, linkcolor=blue, citecolor=blue, urlcolor=blue]{hyperref}
\usepackage{url}            % simple URL typesetting
\usepackage{booktabs}       % professional-quality tables
\usepackage{amsfonts}       % blackboard math symbols
\usepackage{nicefrac}       % compact symbols for 1/2, etc.
\usepackage{microtype}      % microtypography
\usepackage{xcolor}         % colors
\usepackage{fontawesome5}   % GitHub repository icon

\usepackage{siunitx}

\usepackage{amsthm}
\theoremstyle{plain}
\newtheorem{theorem}{Theorem}[section]
\newtheorem{lemma}[theorem]{Lemma}

\newtheorem{corollary}[theorem]{Corollary}

\theoremstyle{definition}
\newtheorem{assumption}[theorem]{Assumption}

\newtheorem{example}[theorem]{Example}

\theoremstyle{remark}
\newtheorem{remark}[theorem]{Remark}

\newcommand{\indep}{\perp\!\!\!\perp}

\title{Just Ramp-Up: Debiasing Regression-based Estimator for A/B Tests under Network Interference}

\author{%
	Qianyi Chen, Bo Li \\
    School of Economics and Management, Tsinghua University
}

\begin{document}

\maketitle

\begin{abstract}
	Network interference complicates A/B testing on online platforms, such as social networks and marketplaces, where causal methods based on single experiments often suffer from significant bias due to complex interference patterns. This paper demonstrates the statistical benefits of merging data from multiple experiments with varying treatment proportions. Sequential experimentation with increasing traffic, or ramp-up, is widely used in tech companies for risk management and cost control. Beyond operational benefits, we show that regression-based estimators trained on merged data achieve substantial bias reduction, even under simple randomization schemes and regression models.
    We focus on the global average treatment effect (GATE), a key estimand in the tech industry, and consider a general interference pattern that extends beyond the 1-hop setting. We present a closed-form bias–variance analysis of the linear regression estimator and show that, in practical settings, the bias term dominates. Moreover, we characterize how merging data across ramp-up stages improves regression training and reduces bias. We also offer an intuitive explanation for this reduction and highlight the synergy between cluster-level randomization and our approach.
    Furthermore, we consider a refined estimator based on graph neural networks (GNN). Extensive simulations across challenging scenarios confirm that our methodology significantly improves the accuracy of regression-based estimators. Code is available at \href{https://github.com/Cqyiiii/Just-Ramp-up}{\mbox{\faGithub~\nolinkurl{github.com/Cqyiiii/Just-Ramp-up}}}.
\end{abstract}

% \section{Introduction}

% A/B tests, also known as randomized controlled trials, have been widely adopted by modern online platforms as a paradigm of data-driven decision-making to support rapid product iteration. In most cases, the causal estimand is the GATE, as experimenters aim to measure the improvement in performance metrics—such as user engagement—resulting from introducing a new product feature to all users compared to maintaining the status quo.
% In complex systems like social networks and marketplaces, the classic Stable Unit Treatment Value Assumption (SUTVA) in causal inference is often violated in cases of social interaction and competition, where the outcome of one user is influenced by the treatments of neighboring users. In such scenarios, the challenge of causal inference under interference naturally arises~\cite{sobel2006randomized,rosenbaum2007interference,hudgens2008toward}. This interference, mediated through a graph—such as a social network or market bipartite graph—is called network interference.

\section{Introduction}

A/B testing has become a core paradigm for data-driven decision-making on modern online platforms. In many applications, the target causal estimand is the global average treatment effect (GATE), which measures the difference in average outcomes between globally deploying a new feature and maintaining the status quo. In complex systems such as social networks and marketplaces, however, this task is complicated by network interference: a unit's outcome may depend not only on its own treatment, but also on the treatments assigned to other connected units~\cite{sobel2006randomized,rosenbaum2007interference,hudgens2008toward}.

Estimating GATE under network interference is particularly challenging in practice because platforms rarely expose all users to treatment during experimentation. Instead, new features are typically rolled out gradually through a ramp-up process, starting from a very small treatment proportion and increasing traffic only for promising variants~\cite{xu2018sqr,kohavi2020trustworthy}. As a result, the available data are collected under limited treatment proportions, whereas GATE concerns two extreme counterfactual regimes: global treatment and global control. This makes GATE estimation fundamentally an extrapolation problem, and interference further amplifies its difficulty~\cite{parker2017optimal}.

% In this paper, we show that the ramp-up process provides not only operational benefits but also a valuable statistical opportunity. We propose to merge data from multiple experiments conducted at different ramp-up stages and use the pooled data to train a regression model for counterfactual prediction. 
% The key advantage of this merging procedure is that it enriches the variation in units' treatment exposures (i.e., the summary statistics of interference on a specific unit, e.g., the proportion of treated ones among a unit's one-hop neighbors on network), thereby helping the regression model generalize to the regime of global treatment.

% In this paper, we show that the ramp-up process provides not only operational benefits but also a valuable statistical opportunity. We propose to merge data from experiments conducted at different ramp-up stages and use the pooled data to train a regression model for counterfactual prediction. The key advantage of this merging procedure is that it enriches the variation in units' treatment exposures, i.e., summary statistics of interference on a given unit, such as the proportion of treated neighbors in its 1-hop neighborhood, thereby helping the regression model generalize to the regime of global treatment.
The ramp-up process also provides a statistical opportunity: data from different stages can be pooled to train a regression model for counterfactual prediction. Pooling enriches variation in units' treatment exposures---summaries of interference such as the fraction of treated 1-hop neighbors---and helps the model extrapolate to global treatment.

Our focus is on regression-based estimators for static GATE. Beyond their practical advantages, such as ease of implementation and the ability to incorporate covariates~\cite{gao2025causal}, regression-based estimators are especially attractive here because they can accommodate substantially richer interference structures than traditional estimators without a regression component. Much of the existing literature relies on neighborhood interference or related low-dimensional assumptions, which reduce the effect of the full treatment vector on a unit to the treatments within its 1-hop neighborhood~\cite{forastiere2022estimatingbayesian,cortez2022agnostic,ugander2023randomized,liu2024cluster,kandiros2026conflict,chen2026journey}. While such assumptions offer important technical simplifications, spillover beyond the 1-hop neighborhood remains underexplored despite its practical relevance. 
To address this gap, we study a novel interference pattern, \textit{general linear interference}, which captures multihop spillover while still permitting \textit{closed-form} bias and variance analysis, in contrast to the classical approximate neighborhood interference framework~\cite{leung2022ani}, which puts bounded-difference control on the impact of long-distant neighbors.
Our analysis also incorporates the network topology explicitly, rather than relying only on coarse graph summaries such as the maximum degree, as is common in the literature.

We also emphasize the role of multiple experiments. While ramp-up experiments are standard in industry, most existing methods for network interference focus on a single experiment. There are some notable exceptions. For example, \cite{cortez2022agnostic} models the mean outcome as a one-dimensional function of the treatment proportion and extrapolates to global treatment and control, while \cite{modeling23namkoong} uses multiple experiments for model selection over candidate interference structures. By contrast, our contribution is to show that multiple experiments can be pooled to improve regression-based GATE estimation itself. We interpret GATE estimation as a counterfactual prediction problem between two extreme exposure regimes, and the central benefit of merging lies in improving the training of the regression function through \textbf{richer network exposure variation} among units.

Our theoretical analysis starts with a simple network-agnostic linear regression estimator under general linear interference. Despite the complexity of long-range spillovers, this model admits a \textit{closed-form} bias--variance decomposition and yields insights into more expressive regression-based estimators. We show that bias dominates estimation error in practically relevant regimes with regular interference intensity. Merging data across ramp-up stages substantially reduces this bias by varying treatment proportions and enriching the resulting network exposures.
% We further characterize how merging data across ramp-up stages can substantially reduce this bias: increasing variation in treatment proportions and resulting units' exposures.

Motivated by this finding, we then study a more expressive estimator based on graph neural networks (GNNs). Extensive simulations show that the same qualitative pattern persists: merged data markedly improve estimation accuracy, and powerful models such as GNNs benefit especially strongly from the richer exposure variation induced by merging. In this paper, we intentionally examine the performance of lightweight models, in contrast to GNN-based approaches with carefully designed architectures~\cite{wu2025causal,huang2026modeling}. Our key message is that, with our data-centric ramp-up augmentation, even simple regression models can achieve strong performance in GATE estimation.

\begin{figure}
    \centering
    \begin{minipage}{0.48\textwidth}
      \centering
      \includegraphics[width=\linewidth]{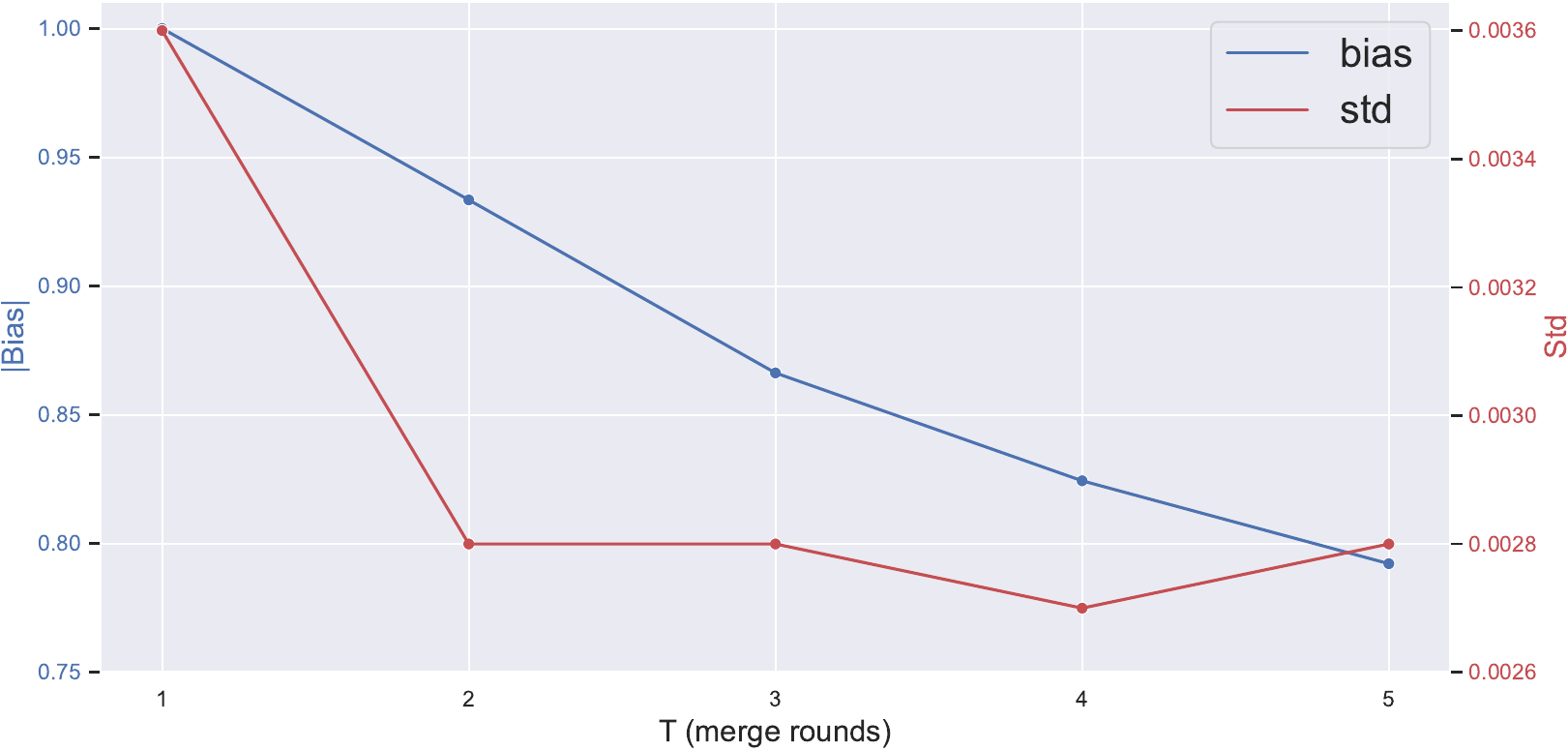}
      \subcaption{linear regression}
    \end{minipage}\hfill
    \begin{minipage}{0.48\textwidth}
      \centering
      \includegraphics[width=\linewidth]{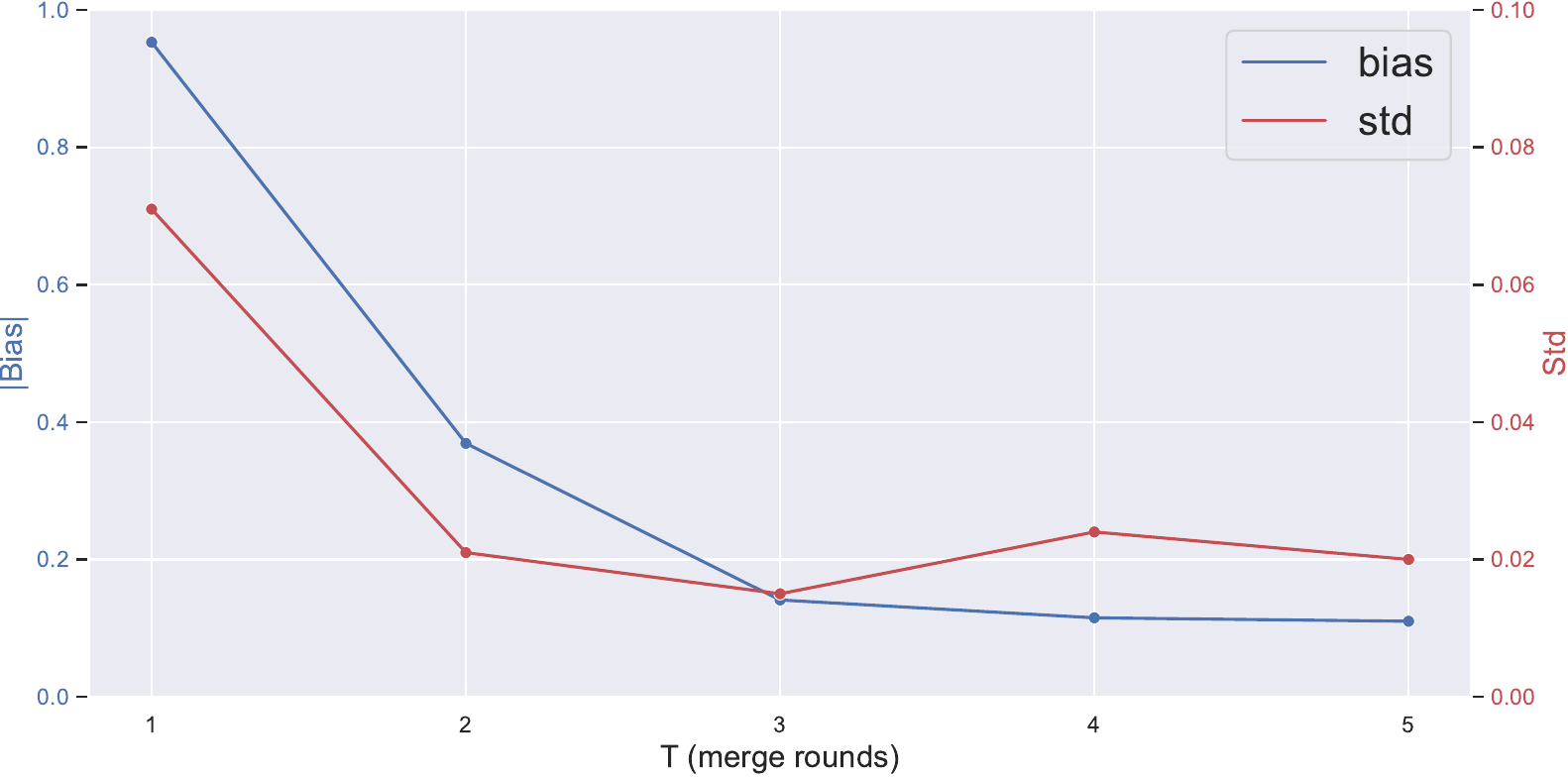}
      \subcaption{graph neural network}
    \end{minipage}
    \caption{Absolute bias (blue curve, left axis) and standard deviation (red curve, right axis) of (a) linear regression and (b) GNN estimators trained on \textbf{merged} ramp-up data. The horizontal axis shows the number $T$ of merged stages. At each $T$, training uses the final $T$ stages, $(c_{6-T},\dots,c_5)$, where $(c_1,\dots,c_5)=(2\%,5\%,10\%,25\%,50\%)$. Thus, $T=1$ uses only the final experiment at 50\% treatment, while $T=5$ pools all five stages.
	% The randomization scheme used here is unit-level complete randomization.
	}\label{fig:combined}
\end{figure}

Figure~\ref{fig:combined} provides an illustrative example. For both linear regression and GNN estimators, merging data from multiple ramp-up stages leads to a clear reduction in bias, while the variance remains comparatively small. The figure also highlights a key practical message: even expressive models such as GNNs perform unsatisfactorily when trained on a single experiment, but become much more effective once trained on merged datasets with distinct treatment proportions. This comparison changes both sample size and treatment-proportion diversity; the repeated-experiment control in Section~\ref{ch5} examines whether additional observations at a fixed proportion offer the same benefit.

Our main takeaway is therefore simple: by pooling ramp-up experiments with different treatment proportions, one can train a regression model that more accurately predicts potential outcomes under global treatment and global control. Since ramp-up is already standard practice on many online platforms, this methodology requires no additional experimentation cost and can be deployed as a plug-and-play improvement for GATE estimation under network interference.

\section{Basic Setting}\label{ch2}

\subsection{Preliminary}

We adopt the Neyman-Rubin potential outcome framework~\cite{rubin1974estimating} and consider a finite population of $n$ units. The treatment vector is represented as $\mathbf{z} = \left(z_1, z_2, \ldots, z_n\right) \in \{0,1\}^n$. Without assuming the Stable Unit Treatment Value Assumption (SUTVA), we define the potential outcome of unit $i$ as $Y_i = Y_i(\mathbf{z})$ to account for its possible dependency on the treatments of other units. The parameter of interest is the global average treatment effect (GATE), which is defined as:
\begin{equation}
    \tau\coloneqq\frac{1}{n} \sum_{i \in[n]}\left(Y_i(\mathbf{1})-Y_i(\mathbf{0})\right) .
\end{equation}
Here, $[n]=\{1,2,\dots,n\}$ is the unit set, and $\mathbf{1}$ and $\mathbf{0}$ are the $n$-dimensional all-ones and all-zeros vectors, representing global treatment and global control. We take the network as a given undirected graph $\mathcal{G}=(\mathcal{V},\mathcal{E})$ with node set $\mathcal{V}=[n]$ and adjacency matrix $A$. Let $\deg_i$ denote the degree of node $i$ and write $D=\operatorname{diag}(\deg_1,\dots,\deg_n)$. We assume every node has positive degree.

\subsection{General Linear Interference Model}

We consider the following data generation mechanism, called \textit{general linear interference}:
\begin{equation}\label{pom}
    Y(\mathbf{z}) = \beta_0\mathbf{1} + \beta_1 \mathbf{z} + B\mathbf{z} + \boldsymbol{\epsilon},
\end{equation}
where $(\beta_0,\beta_1)$ and $B$ are all unknown parameters. With these, the true GATE is given by:
\begin{equation}
    \tau = \beta_1 + \frac{1}{n}\mathbf{1}^\top B \mathbf{1},
\end{equation}
For convenience, we further assume an i.i.d. mean-zero noise structure.
\begin{equation}
    % \epsilon\sim {N}(\mathbf{0}, \sigma_e^2I_n)  \qquad     \mathbf{z} \indep \mathbf{\epsilon}
    \mathbb{E}[\boldsymbol{\epsilon}]=\mathbf{0}, \qquad \operatorname{Var}(\boldsymbol{\epsilon})=\sigma_e^2 I_n, \qquad \mathbf{z} \indep \boldsymbol{\epsilon}.
\end{equation}
The matrix $B$ in Equation (\ref{pom}) characterizes interference effects. In this paper, we impose $B_{ii}=0$ for all $i\in[n]$ to separate the direct treatment effect from the interference effect. For simplicity, we do not model heterogeneity in the baseline effect $\beta_0$ or direct effect $\beta_1$ across units. Instead, we place the modeling flexibility on a general interference matrix $B$, which can capture complex and potentially long-range dependencies on the network. We provide two examples of such matrices $B$ below.

\begin{example}[1-hop linear potential outcome model]\label{example1}
    If we take $B = r D^{-1}A$, where $r$ is the scale parameter, then our potential outcome model is reduced to the classic linear-in-means model~\cite{aronow2017estimating}:
    \begin{equation}
 Y_i(\mathbf{z}) = \beta_0 + \beta_1 z_i + r \frac{\sum_{j\in \mathcal{N}(i)} z_j}{\deg_i} + \mathbf{\epsilon}_i,
    \end{equation}
    where $\mathcal{N}(i)$ denotes the 1-hop neighborhood of unit $i$ in the graph.
\end{example}

% We then provide an example that incorporates multihop interference, which demonstrates the generality of this potential outcome model.
\begin{example}[Multihop linear potential outcome model]\label{example2}
    We consider
    \begin{equation}
        B = (\mathbf{1}\mathbf{1}^\top - I_n) \odot \left(\sum_{m=1}^M r_m (D^{-1}A)^m\right),
    \end{equation}
    where $\odot$ denotes the Hadamard product, $M$ is the interference range, and $r_1 \geq r_2 \geq \cdots \geq r_M > 0$.
\end{example}

Notice that $A^m$ characterizes the $m$-hop connections on the graph, where $A^m_{ij}$ represents the number of walks of length $m$ connecting units $i$ and $j$. The term $D^{-1}$ is included here as a normalization factor. The sequence of scale parameters, $\{r_i\}_{i=1}^M$, describes how interference decreases with increasing distance on the graph. Additionally, we exclude the diagonal elements of matrix $B$. This exclusion addresses the identification issue and can also be interpreted as eliminating the circular effects associated with closed walks that begin and end at the same unit.
Since $D = \operatorname{diag}(A\mathbf{1})$, we can summarize the examples above as $B = g(A)$ for certain function $g$, namely, $B$ models the interference that is conducted through the network topology. We consider the proposed general linear model for carrying out \textit{closed-form} bias and variance analysis while maintaining sufficient generality.

\subsection{Experimental Design}

In this section, we discuss our default experimental design. Specifically, we adopt unit-level \emph{complete randomization} as the basic randomization scheme, which randomly chooses exactly $d$ units from a population of n units to receive treatment. To formalize it, we uniformly sample the binary treatment vector $\mathbf z$ from $\left\{\mathbf z\in\{0,1\}^n:\sum_{i=1}^n z_i=d\right\}$. In particular, $\mathbb E[z_i]=d/n$.

% satisfying $\sum_{i=1}^n z_i = d$. In particular, $\mathbb{E}[z_i]=d/n$.

% We adopt unit-level \emph{complete randomization}: the treatment vector
% $\mathbf z$ is drawn uniformly from
% \[
% \mathcal Z_d := \left\{\mathbf z\in\{0,1\}^n:\sum_{i=1}^n z_i=d\right\}.
% \]
% Equivalently, exactly $d$ units are treated and each size-$d$ treatment set
% is equally likely. In particular, $\mathbb E[z_i]=d/n$.

% we restrict binary treatments $\{z_i\}_{i=1}^n$ to satisfy $\sum_{i=1}^{n} z_i = d$ besides $\mathbb{E}[z_i]=d/n$ in a single experiment. 

We note that the performance of unit-level complete randomization is often very similar to the independent Bernoulli randomization, which assigns treatment to each unit with probability $p=d/n$ independently. However, unit-level complete randomization ensures exactly $d$ treated units, which brings considerable convenience in our derivation for closed-form expressions of bias and variance.

We next discuss the setting of multiple experiments. We consider a ramp-up process that includes $T$ experiments that are subject to complete randomization with treated units $d_1<d_2<\dots<d_T$. For convenience, we introduce another notation $c_t = d_t/n$. For instance, $T=5$ and $(c_t)_{t=1}^T   = (2\%, 5\%, 10\%, 25\%, 50\%)$, as also illustrated in~\cite{xu2018sqr}. Since our core methodology is merging multiple experimental data, we will also discuss the selection of $T$ and $(c_t)_{t=1}^T$ later in the analysis.

\subsection{Regression-based Estimator}

We consider a regression-based estimator based on a function $f \in \mathscr{F}$ for predicting the counterfactual $Y(\mathbf{z})$, where $f: \{0,1\}^n \times \mathscr{A} \to \mathbb{R}^n$ and $\mathscr{A}$ is the space of adjacency matrices of simple graphs on $n$ nodes. Given a prediction function $f$ trained on experimental data, we define the estimator as the difference between the predicted mean outcomes under global treatment and global control:
\begin{equation}
 \hat \tau(f) = \frac{1}{n} \mathbf{1}^\top \left( f(\mathbf{1}, A) - f(\mathbf{0}, A)\right).
\end{equation}
In classical causal inference, the augmented inverse propensity weighting (AIPW; \cite{chernozhukov2018double}) estimator is generally preferred to purely regression-based estimators. Under network interference, however, inverse propensity weights often lead to explosive variance~\cite{ugander2023randomized,chen2026journey} and are also vulnerable to exposure misspecification. Regression-based estimators therefore occupy a distinctive and important role. In this paper, we focus on two regression model classes, linear regression $\mathscr{F}_{\text{LR}}$ and graph neural networks $\mathscr{F}_{\text{GNN}}$, and develop a data-centric approach for bias reduction in regression-based estimation.

% The regression function is trained through empirical risk minimization on the experimental data collected so far. In the next section, we will discuss how to construct better training data to unleash the potential of regression-based estimators from a statistical perspective. Additionally, $f \in \mathscr{F}_{\text{GNN}}$ is well-suited to scenarios where the graph structure is dynamic—such as when the network topology is updated weekly on online platforms. It is entirely feasible to train a regression model on merged data with varying network topologies, e.g., ${(\mathbf{z}_1, A_1), (\mathbf{z}_2, A_2)}$.

\section{Analysis on Linear Regression Estimator}\label{ch3}

We analyze a graph-agnostic, deliberately misspecified linear regression model under general linear interference. Its tractable bias and variance illustrate how merging data affects training. Our simulations show that these qualitative insights also apply to more expressive GNN models.

We fit $f(\mathbf{z},A)=X(\mathbf{z})\hat\theta$, with $X(\mathbf{z})=(\mathbf{1},\mathbf{z})$ and ordinary least squares (OLS) coefficient
\begin{equation}
    \hat\theta = (X(\mathbf{z})^\top X(\mathbf{z}))^{-1} X(\mathbf{z})^\top Y.    
\end{equation}
We retain the notation $X(\mathbf{z})$ to emphasize its dependence on the experimental design.

Linear regression is technically convenient because it admits a closed-form estimator, unlike many regression-based methods defined via variational formulations. The main difficulty is that the design matrix $X(\mathbf{z})$ is itself random under treatment assignment, so $(X(\mathbf{z})^\top X(\mathbf{z}))^{-1}$ contains a random determinant in the denominator, severely hindering exact analysis. However, we find that \textit{complete randomization} offers a crucial source of tractability by fixing the number of treated units.

% ch3.1

\subsection{One-step Experiment}

Throughout the section, we use $\hat\tau_{(T)}$ to denote the linear regression estimator based on merged $T$-step experiments. We now examine the case of $T=1$ as a preparation. We specify the number of treated units in the experimental data as $d=cn$ and derive the bias and variance for $\hat\tau_{(1)}$.

\begin{theorem}\label{theorem1}
The bias and variance of $\hat\tau_{(1)}$ under general linear interference are given by:
% {\small
\begin{equation}
    \begin{aligned}\label{eq:1step_bias}
        \operatorname{Bias}(\hat \tau_{(1)}) = -\frac{(\sum_{i,j} B_{ij})}{n-1},
    \end{aligned}
\end{equation}
and 
    \begin{equation}
\begin{aligned}
\operatorname{Var}(\hat \tau_{(1)})
&= \left(\frac{1}{nc(1-c)}\right)^2
\left(
(\sum_{i,j} B_{ij})^2
\left(\frac{c^3(c-1)}{n} + O\left(\frac{1}{n^2}\right)\right) \right.
\\
&\quad
+ (\sum_{i,j,k} B_{ij}B_{kj})
\left(\frac{c(c-1)}{n}+O\left(\frac{1}{n^2}\right)\right)
+ (\sum_{i,j,l} B_{ij}B_{il})
\left(c^3(1-c) + O\left(\frac{1}{n}\right)\right)
\\
&\quad\left.
+ (\sum_{i,j} B_{ij}^2)
\left(c^2(1-c) + O\left(\frac{1}{n}\right)\right)
+ (\sum_{i,j} B_{ij}B_{ji})
\left((1-c)^2c^2 + O\left(\frac{1}{n}\right)\right)
\right)
\\
&\quad
- (\sum_{i,j} B_{ij})^2\frac{1}{(n(n-1))^2}
+ \frac{\sigma_e^2}{nc(1-c)} .
\end{aligned}
\end{equation}
% }
\end{theorem}
Here, notations like $\sum_{i,j}$ denote summation over $i\in[n], j\in[n]$ without additional constraints. 
We suppress lower-order terms for readability; the complete expansion is given in Appendix~\ref{proof_of_theo1}, and the hidden constants are indeed small. 
We now proceed to elaborate on the results presented above. First, we notice that the intensity of direct effect, $\beta_1$, does not influence either bias or variance. In total, the primary source of bias is interference, which corresponds to the second term in the expression.

% The variance seems more complex than the bias. The most intricate part involves the expectation of a squared term, which we expand by categorizing the quadruple $(i, j, k, l)$ scenarios. Specifically, we evaluate the value of the terms $\mathbb{E}[B_{ij}B_{kl} z_i(z_j-c)z_k(z_l-c)]$. This naturally raises a discussion on the matrix $B$, i.e., the summation $\sum_{(i,j,k,l) \in \mathcal{I}}B_{ij}B_{kl}$
% for certain index set $\mathcal{I} \subseteq [n]^4$. For example, $\mathcal{I} = \{(i,j,k,l) \mid  i=k, j=l, (i,j,k,l)\in [n]^4  \}$ corresponds to $\sum_{i,j}B_{ij}^2$ in Theorem~\ref{theorem1}.

% To obtain more insightful results, we choose to control the order of such a complex expression of variance. The key message from the variance analysis is that in the regular case—where interference is comparable to the direct effect—the variance is negligible relative to the (squared) bias in magnitude. To formalize the scenario of interest, we introduce Assumption \ref{assumption1} that limits our analysis to cases of regular interference. Rather than constraining the order of five summations in Theorem \ref{theorem1} separately, we find that it can boil down to three succinct and reasonable conditions.

Although the variance expression involves several quadratic functionals of
$B$, the main takeaway is simple: under a regular interference regime (the interference is comparable to the direct effect), all
of these terms are controlled at a lower order than the squared bias.
This motivates the next assumption, which compresses the relevant
requirements into three interpretable conditions.
\begin{assumption}[Regular interference]\label{assumption1}
With general linear interference, we further assume that 
    \begin{equation}
        \sum_{j=1}^{n} |B_{ij}| = O(1) \qquad \forall i\in[n].
    \end{equation}
Moreover, 
    \begin{equation}
        \sum_{i,j} B_{ij} = \Theta(n) \qquad 
        \sum_{j} (\sum_i B_{ij})^2 = O(n).
    \end{equation}
\end{assumption}
The three conditions serve distinct purposes: the first bounds the total spillover received by each unit, the second ensures that average interference remains non-negligible, and the third rules out a vanishing fraction of highly influential nodes from dominating the variance. Under this regime, the variance is of lower order than the squared bias. We further provide a detailed discussion on this regime in Appendix~\ref{app:discuss_assump1}. Under Assumption~\ref{assumption1}, we obtain the following corollary:
\begin{corollary}\label{corollary1}
    Based on Assumption~\ref{assumption1} and Theorem~\ref{theorem1}, we further conclude that 
    \begin{equation}    
            \operatorname{Bias}(\hat \tau_{(1)}) = \Theta(1) \qquad
            \operatorname{Var}(\hat \tau_{(1)}) = \Theta(1/n).
    \end{equation}
\end{corollary}
This corollary shows that, for the estimator and regime considered here, the main challenge introduced by interference lies in the bias rather than the variance, especially since social networks on online platforms typically involve a large number of units. Therefore, to reduce the mean-squared error (MSE) of the estimator effectively, our primary focus should be on addressing the bias term.

\subsection{Two-step Experiment}

Although the derivation for $T>2$-step experiments does not introduce substantial additional complexity compared with the case of $T=2$, we discuss the latter separately because the benefit of merging experimental data from two ramp-up stages with distinct treatment proportions is fundamental, compared to the single-experiment case.
First, we outline the changes induced by merging two experimental datasets. 
For merged data, we stack the observations from different ramp-up stages and
fit one OLS regression on the resulting $2n$ rows. Thus each stage contributes
$n$ observations, while the treatment proportions differ across stages. In the
baseline analysis, the stage-specific noises are independent across stages and
units, with common variance $\sigma_e^2$. The treatment proportions in two experiments are $c_1$ and $c_2$. Moreover, there are two additional changes:
\begin{equation}\label{eq:expand}
    \begin{aligned}
        B\xrightarrow{\text {expand}}
        \begin{pmatrix}
            B & 0\\ 0& B    
        \end{pmatrix}
        \qquad 
        \mathbf{z} \xrightarrow{\text {expand}}
        \begin{pmatrix}
            \mathbf{z}_1 \\ \mathbf{z}_2
        \end{pmatrix}.
    \end{aligned}
\end{equation}

% \begin{remark}
%     In the transition from $T=1$ to $T=2$, it can be demonstrated that our analytical framework is \textbf{flexible enough to incorporate temporal interference and dynamic graph structures}. For example, the upper-triangular block of the matrix $\operatorname{diag}(B, B)$ can capture general forms of temporal interference—specifically, cases where the treatment assignment of unit $i$ at first round can linearly affect the outcome of unit $j$ at second round. A minor modification to the framework can also be made to control temporal interference when its influence is limited in scale. Nevertheless, \textbf{incorporating temporal interference alters the estimand} and complicates the analysis, while we \textbf{focus on the cross-sectional component} in this paper. Regarding the dynamic graph structure, the case just becomes $\operatorname{diag}(B_1, B_2)$, as we assume the interference matrix $B$ is transformed from the adjacency matrix $A$. We further examine the case of dynamic graph structure in our simulation study.
% \end{remark}
\begin{remark}
%     The block-diagonal expansion in~\eqref{eq:expand} rules out cross-round spillovers because we focus on the cross-sectional part (see also~\cite{modeling23namkoong}) and estimating the static GATE. Actually, temporal interference can be incorporated by replacing $\operatorname{diag}(B,B)$ with
% \[
%     \begin{pmatrix}
%         B_1 & 0\\
%         B^{(1\to 2)} & B_2
%     \end{pmatrix},
% \]
% where $B^{(1\to 2)}$ captures the linear effect of first-round treatments on second-round outcomes. Dynamic graphs can likewise be accommodated by allowing $B_1 \neq B_2$. We focus on the block-diagonal case because temporal interference changes the estimand and would blur the cross-sectional bias structure of interest. Nevertheless, our framework is flexible enough to incorporate both temporal issues.
The block-diagonal expansion in~\eqref{eq:expand} rules out cross-round spillovers, as our focus is on the cross-sectional interference (see also~\cite{modeling23namkoong}) and on estimating the static GATE. In principle, temporal interference can be incorporated by replacing $\operatorname{diag}(B,B)$ with
    $\begin{pmatrix}
        B_1 & 0\\
        B^{(1\to 2)} & B_2
    \end{pmatrix}$,
where $B^{(1\to 2)}$ captures the linear effect of first-round treatments on second-round outcomes. Dynamic graphs can likewise be accommodated by allowing $B_1 \neq B_2$. We focus on the block-diagonal case because temporal interference changes the estimand and would blur the cross-sectional bias structure of interest. Nevertheless, our framework is flexible enough to accommodate both temporal extensions.
\end{remark}
We then characterize the bias of the 2-step estimator $\hat\tau_{(2)}$ in the following theorem. In the next subsection, we will show that the variance for general $\hat{\tau}_{(T)}$ remains negligible compared to the bias under Assumption~\ref{assumption1}. We maintain the setting $c_1,c_2\in(0,1)$ and $c_1<c_2$ as introduced in Section~\ref{ch2}. 
\begin{theorem}\label{theorem2}
    The bias of the $\hat\tau_{(2)}$ under general linear interference is given by:    
    \begin{equation}
        \begin{aligned}
            \operatorname{Bias}(\hat \tau_{(2)}) &= 
            -
            \frac{\sum_{ij}B_{ij}}{n} \left( 1 - \frac{(n-1)(c_1-c_2)^2+2(c_1^2+c_2^2)-2(c_1+c_2)}{(n-1)(c_1+c_2)(2-(c_1+c_2))} \right).
        \end{aligned}
    \end{equation}
When $n>2(c_1+c_2-c_1^2-c_2^2)/(c_1-c_2)^2 +1$, we have:
\begin{equation}
    % \frac{(n-1)(c_1-c_2)^2+2(c_1^2+c_2^2)-2(c_1+c_2)}{(n-1)(c_1+c_2)(2-(c_1+c_2))} \in (0,1)
	0<\frac{(n-1)(c_1-c_2)^2+2(c_1^2+c_2^2)-2(c_1+c_2)}{(n-1)(c_1+c_2)(2-(c_1+c_2))} < 1.
\end{equation}
This directly implies that $\operatorname{Bias}(\hat\tau_{(1)})$ and $\operatorname{Bias}(\hat\tau_{(2)})$ share the same sign. A substantial reduction in the magnitude of bias from $\hat\tau_{(1)}$ to $\hat\tau_{(2)}$ is given by:
\begin{equation}
    |\operatorname{Bias}(\hat\tau_{(1)})|-|\operatorname{Bias}(\hat\tau_{(2)})|=\frac{|\sum_{ij}B_{ij}|}{n} 
    \frac{(c_1-c_2)^2}{(c_1+c_2)(2-(c_1+c_2))} + O\left(\frac{1}{n}\right).
\end{equation}
\end{theorem}

From the perspective of the online platform, $c_1+c_2$ can be viewed as the total budget to be allocated in the experiment. If we fix $c_1 + c_2$, increasing $|c_1 - c_2|$ corresponds to an increase in exposure variation and a decrease in bias. 
This bias reduction is intuitive, since estimating the GATE inherently requires estimating the two mean outcomes under treatment proportions $0$ and $1$, and lowering $c_1$ while increasing $c_2$ brings us closer to these target regimes, reflecting a similar intuition in~\cite{eckles2016design}, namely that `moving closer to the counterfactual situation of interest' induces bias reduction.

% BEGIN MOVED FROM APPENDIX: A2_TRAFFIC_CAP
A complementary constraint caps the treatment proportion at $\max\{c_1,c_2\}\leq c$ to limit risk to revenue and user experience. Setting $c_1=0$ and $c_2=c$ maximizes bias reduction under this cap, yielding
\begin{equation}\label{eq:traffic_cap_bias}
    -\frac{\sum_{ij}B_{ij}}{n}\left(1 - \frac{c}{2-c}\right) + O\left(\frac{1}{n}\right),
\end{equation}
which is still numerically unsatisfactory, particularly since $c \leq 0.5$ is common in practice.
% END MOVED FROM APPENDIX: A2_TRAFFIC_CAP

Moreover, somewhat counterintuitively, merging \textit{repeated} experiments with the same treatment proportion $c$ does \textbf{not} substantially reduce bias, even though it increases the amount of data drawn from the same distribution. We discuss the implications of Theorem~\ref{theorem2} further in Appendix~\ref{app:discuss_theorem2}. Finally, since the variance analysis and conclusions for $T>2$ closely parallel those for $T=2$, we defer the detailed results to the next section. In particular, the variance remains of order $\Theta(1/n)$ and is therefore still negligible compared to the bias when merging $T$ ramp-up stages.

\subsection{\texorpdfstring{$T$}{}-step Experiment}
In this section, we show that merging $T$ ramp-up stages can still substantially reduce bias, although adding more intermediate steps may even be harmful compared to merging only the lowest and largest treatment proportions. We then establish that the variance remains secondary relative to the bias. Here, we emphasize that our regime is one in which $n$ is large while $T$ remains a small fixed constant (e.g., $n=10000$ and $T=4$), in line with practical settings on technology platforms.
\begin{theorem}\label{theorem3}
    Assume $T$ is fixed. The bias of $\hat\tau_{(T)}$ under general linear interference is given by: 
    \begin{equation}
        \begin{aligned}
            \operatorname{Bias}(\hat \tau_{(T)})&= 
            -
            \frac{\sum_{ij}B_{ij}}{n} 
            \left( 1 - \frac{T\sum_{t=1}^T c_t^2 - (\sum_{t=1}^T c_t)^2}{(\sum_{t=1}^T c_t)(T - \sum_{t=1}^T c_t)} \right) + O\left(\frac{1}{n}\right).
        \end{aligned}
    \end{equation}
\end{theorem}
Thus, bias reduction depends on the dispersion of treatment proportions across stages. Repeated experiments with the same $c$ have no such dispersion and do not reduce the leading bias.
% Using the same argument as in Theorem~\ref{theorem1} together with the Cauchy--Schwarz inequality, we can verify that merging data from a $T$-step experiment with varying treatment proportions still yields substantial bias reduction compared to using a single experiment. 
Moreover, simple algebra shows that, for fixed $c_1,\ldots,c_T$, merging more experimental stages does \textbf{not} necessarily lead to further bias reduction, and merging only the two extreme steps $c_1$ and $c_T$ is already sufficiently effective. 
We defer the detailed discussion to Appendix~\ref{app:discuss_theorem3} and Appendix~\ref{app:exp_results}.

Next, we analyze the variance of the regression-based estimator trained on merged data. Merging data across time introduces an additional temporal dimension in the design and may in principle create new covariance components. Nevertheless, we show that the variance remains of order $\Theta(1/n)$. We focus on two representative settings: temporally independent experiments and staggered rollout experiments~\cite{xiong2023optimal}, where treatment levels of units are non-decreasing over time.
\begin{theorem}\label{theorem4}
    For the linear regression estimator trained on merged $T$-step data from either \textbf{temporally independent} or \textbf{staggered rollout} experiments, the order of variance is given by: 
    \begin{equation}        
            \operatorname{Var}(\hat \tau_{(T)}) =  \Theta\left({1}/{n}\right).
    \end{equation}
\end{theorem}

The final part of Assumption~\ref{assumption1} restricts the order of the sum of the total impact that unit $i$ imposes on other units. Intuitively, it restricts the number and influence of highly influential opinion leaders within the network. In Example~\ref{example1}, this assumption does not always hold and thus imposes substantial constraints on the adjacency matrix $A$, specifically, $\mathbf{1}^\top D^{-1}A^2 D^{-1}\mathbf{1} = O(n)$.

For instance, in a ring graph, we have $\sum_{j} (\sum_i B_{ij})^2 = \Theta(n)$, while in a star graph—where one opinion leader connects to all other units—we have $\sum_{j} (\sum_i B_{ij})^2 = \Theta(n^2)$.
In the context of Example~\ref{example1}, a sufficient condition for satisfying this assumption is the restricted growth rate that is commonly employed in related works~\cite{ugander2013graph,ugander2023randomized,jia2025clustered}.
However, a restricted growth rate is not a necessary condition; e.g. $\mathbf{1}^\top D^{-1}A^2 D^{-1}\mathbf{1} = n$ in the complete graph; despite being the densest graph, it lacks any prominent opinion leaders.

Lastly, the final part of Assumption~\ref{assumption1} is not required for deriving the Corollary~\ref{corollary1} but is instead intended to support the conclusion in the case of merging data, specifically that the variance of the linear regression estimator remains at the order of $\Theta(1/n)$.
% END MOVED FROM APPENDIX: A1_GRAPH_STRUCTURE_AND_VARIANCE

\section{An Intuitive Perspective on Exposure Variation}\label{ch4}

Although we study the graph-agnostic linear regression estimator for theoretical tractability, the conclusion of the previous section should extend more broadly. The key reason is that merging data across stages increases variation in macro-level treatment proportions, which \textbf{translates to variation in micro-level network exposures}. This added exposure diversity can likewise facilitate learning the interference pattern for more flexible models, such as GNN-based regressors.

\begin{figure}
\centering
\begin{subfigure}{0.32\columnwidth}
    \centering
    \includegraphics[width=\columnwidth]{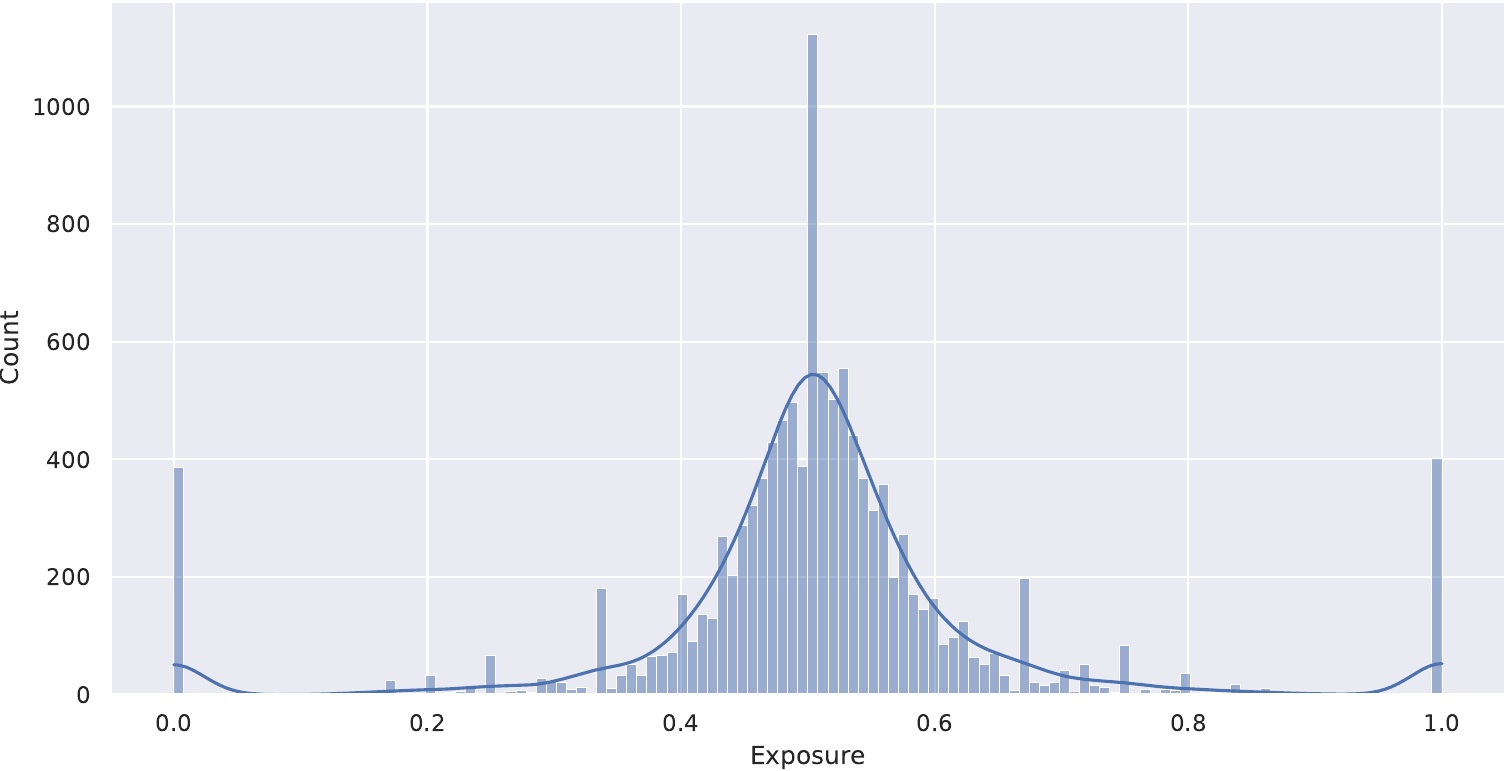}
    \caption{}
    \label{fig:exposure_dist_half}
\end{subfigure}
\hfill
\begin{subfigure}{0.32\columnwidth}
    \centering
    \includegraphics[width=\columnwidth]{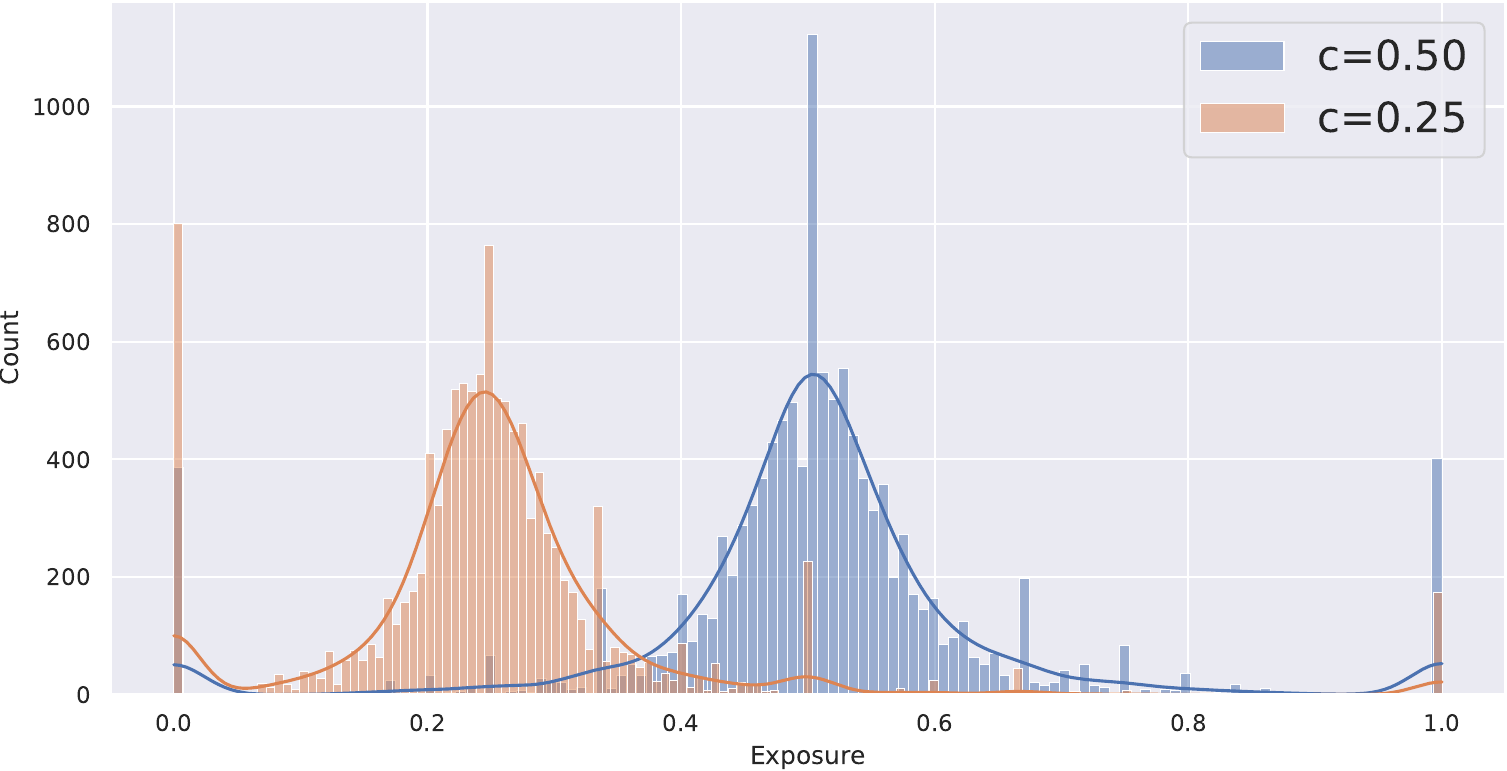}
    \caption{}
    \label{fig:exposure_dist_merge}
\end{subfigure}
\hfill
\begin{subfigure}{0.32\columnwidth}
    \centering
    \includegraphics[width=\columnwidth]{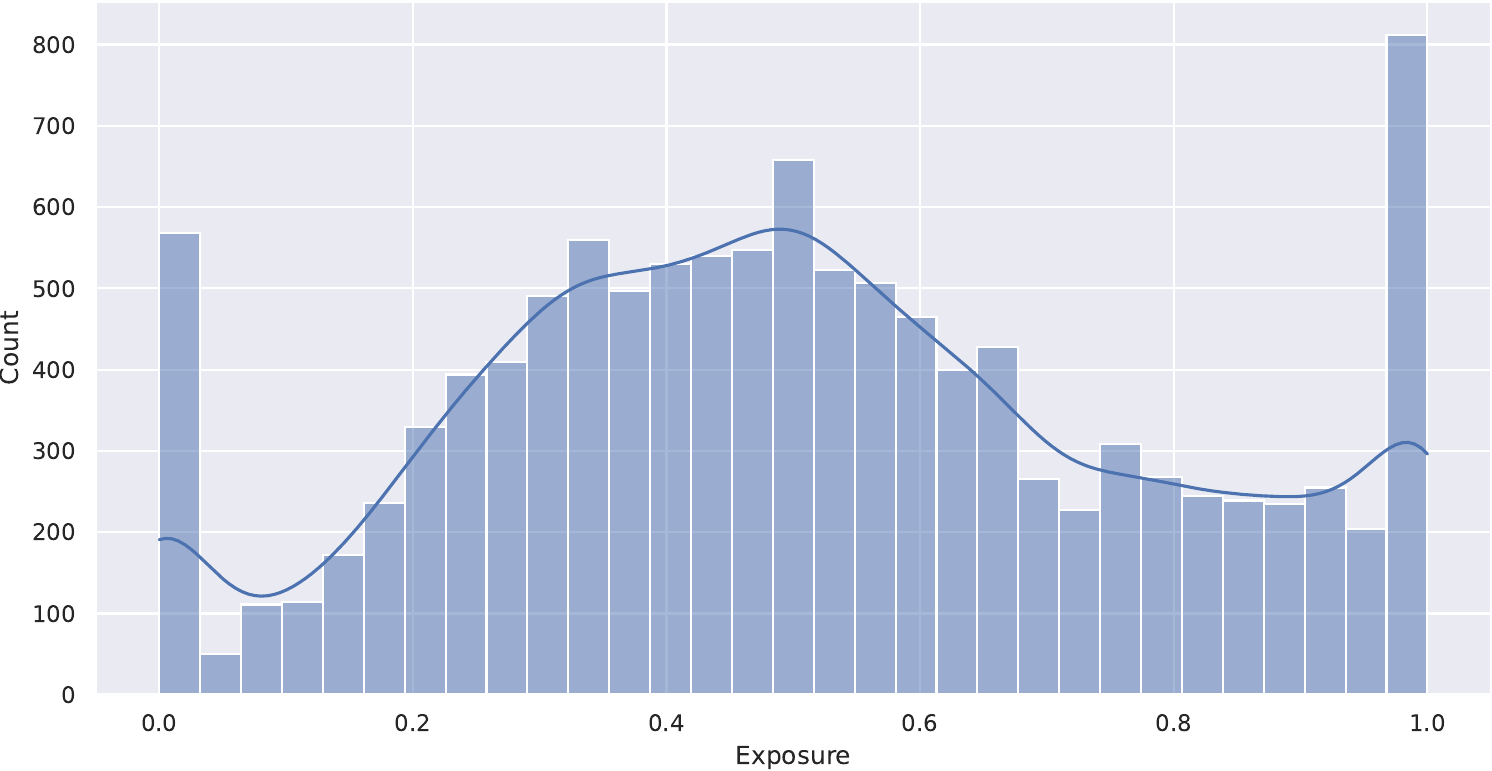}
    \caption{}
    \label{fig:exposure_dist_cluster}
\end{subfigure}
\caption{
Fig.~\ref{fig:exposure_dist_half} shows the exposure distribution under treatment proportion $c=0.5$. Fig.~\ref{fig:exposure_dist_merge} shows the corresponding distribution with $c=0.25, 0.5$, and Fig.~\ref{fig:exposure_dist_cluster} shows that under \textit{cluster-level randomization} with $c=0.5$. The exposure variances are $0.0243$, $0.0377$, and $0.0712$, respectively.}
\label{fig:semi_synthetic}
\end{figure}

Specifically, we use the case $T=2$ and Example~\ref{example1} for illustration. In this classical setting, the treatment exposure of unit $i$ is defined as the fraction of its treated neighbors,
% $e_i = \frac{\sum_{j \in \mathcal{N}(i)} z_j}{\operatorname{deg}_i}.$
$e_i = {\sum_{j \in \mathcal{N}(i)} z_j}/{\deg_i}.$
We argue that the main challenge in modeling interference lies in obtaining sufficient variation in the treatment exposures $\{e_i\}_{i=1}^n$. Fortunately, because each $e_i$ is induced by treatment randomization, the exposures naturally exhibit variation. We visualize their distribution in Figure~\ref{fig:semi_synthetic}.

% A closer examination of the transition from the case of $T=1$ to $T=2$ reveals that the bias reduction stems from the transformation of the term $(\mathbf{z} - c \mathbf{1})$ in the bias expression into $(\mathbf{z}_1 - \frac{c_1 + c_2}{2} \mathbf{1})$ and $(\mathbf{z}_2 - \frac{c_1 + c_2}{2} \mathbf{1})$. Mathematically, the former term has a zero mean, introducing a lower-order component compared to the constant term in bias, whereas the latter terms do not, thereby mitigating the bias. This raises a natural question: what is the underlying intuition behind this improvement beyond mere algebraic manipulation? In this section, we aim to provide insight into the benefits of merging datasets. Specifically, we argue that the merging procedure enhances the variation in treatment exposures.

% In line with the principles of statistical learning, successful model training relies on sufficient variation in the feature space. We note that increasing the variation in covariates $X$ helps reduce the variance of the OLS estimator, $\sigma^2 (X^\top X)^{-1}$, which represents the classical case where unbiasedness is achieved. However, as shown in Section \ref{ch3}, under regular interference, the primary challenge lies in the bias term. Generally, greater variation in data can also reduce the bias of the machine learning model. For example, in Ridge regression, increasing data variation can reduce both variance and bias. 

In a single experiment under complete randomization, $\deg_i e_i$ follows
a hypergeometric distribution and is well-approximated by
$\mathrm{Binom}(\deg_i,c)$ when $n$ is large. The overall exposure
distribution is therefore a mixture of these unit-level laws across $i$, as shown in Figure~\ref{fig:exposure_dist_half}. 
Next, Figure~\ref{fig:exposure_dist_merge} presents two histograms for the case $c_1 = 0.25, c_2 = 0.5$. The variation in treatment exposures within the merged data increases substantially. Beyond the variation induced by randomization, merging data from experiments with different treatment proportions naturally further enriches exposure variation.

Moreover, we note that the possible harmfulness of merging intermediate steps besides the first and $T$-th step, as detailed in Appendix~\ref{app:discuss_theorem3}, can also be explained by this principle, i.e., merging intermediate steps would in turn decrease exposure variation.
% BEGIN MOVED FROM APPENDIX: A3_EXPOSURE_VARIATION
For example, merging stages with treatment proportions $(10\%,50\%)$ yields an exposure variance of 0.0587. Adding a stage with $c=0.25$ reduces this variance to 0.0459, illustrating how an intermediate stage can increase bias. Exposure variance is maximized when half the units have exposure 0 and half have exposure 1. This ideal, generally impractical configuration directly covers both counterfactual regimes defining GATE.
% END MOVED FROM APPENDIX: A3_EXPOSURE_VARIATION

Figure~\ref{fig:exposure_dist_cluster} illustrates cluster-level randomization~\cite{hudgens2008toward,ugander2013graph,eckles2016design}, which assigns a common treatment to units within each network community. These correlated assignments increase exposure variation relative to unit-level randomization: here, the exposure variance rises to $0.0712$. This offers a learning-based explanation for the benefits of cluster randomization. By inducing covariance among treatments, it enriches the exposures available to the regression model. Merging and cluster randomization can therefore act jointly to improve estimation, as the simulations in Figure~\ref{fig:total_results} demonstrate.

Finally, although the illustration is based on a particular exposure definition, we believe this intuition extends naturally to the most general setting, where $e_i = h_i(\mathbf{z}, A)$ with a unit-specific function $h_i$.

% lienar regression also benefits from cluster-level randomization

% \begin{figure}
%     \centering
%     \begin{minipage}{0.48\textwidth}
%       \centering
%       \includegraphics[width=\linewidth]{imgs/two-level-LR.pdf}
%       \subcaption{linear regression}
%     %   \label{fig:}
%     \end{minipage}\hfill
%     \begin{minipage}{0.48\textwidth}
%       \centering
%       \includegraphics[width=\linewidth]{imgs/two-level-GNN.pdf}
%       \subcaption{graph neural network}
%     %   \label{fig:sub2}
%     \end{minipage}
%     \caption{The performance of the regression-based estimators with \textit{unit-level} and \textit{cluster-level} complete randomization. The basic setting is the same as that of Figure~\ref{fig:combined}.} 
%     \label{fig:lr_gnn_perf_cluster_ch3}
% \end{figure}

\section{Simulation Study}\label{ch5}

\subsection{Basic Setting}

% \textbf{Data.} We utilize a Facebook social network FB-Stanford3~\cite{nrfbstanf}, which has also been employed in the experiments of~\cite{ugander2023randomized, chen2024optimized}. The network topology $\mathcal{G}$ consists of $|\mathcal{V}|=11586$ nodes and $|\mathcal{E}|=568309$ edges.  Based on the real network, we generate the potential outcomes with following models: linear-in-means~\cite{aronow2017estimating}, multiplicative~\cite{ugander2023randomized}, quadratic~\cite{cortez2022agnostic}, multihop, and linear-in-means with dynamic graph structure. The linear-in-means and multihop model has been introduced in Example~\ref{example1}, \ref{example2}, respectively (we take $M=2$, i.e., 2-hop), and we leave the details of the last dynamic graph structure in Appendix~\ref{app:exp_po_model}. 
% We introduce the multiplicative and quadratic interference as follows:

\textbf{Data.} We use the Facebook social network FB-Stanford3~\cite{nrfbstanf}, which has also been adopted in the experiments of~\cite{ugander2023randomized,chen2024optimized}. The network topology $\mathcal{G}$ contains $|\mathcal{V}| = 11586$ nodes and $|\mathcal{E}| = 568309$ edges. Based on this real network, we generate potential outcomes under the following models: linear-in-means~\cite{aronow2017estimating}, multiplicative~\cite{ugander2023randomized}, quadratic~\cite{cortez2022agnostic}, multihop, and linear-in-means with dynamic graph structure. The linear-in-means and multihop models were introduced in Examples~\ref{example1} and~\ref{example2}, respectively, and we take $M = 2$ (i.e., 2-hop) here. We defer the details of the dynamic-graph setting to Appendix~\ref{app:exp_setting}. We introduce the multiplicative and quadratic interference models below:
\begin{gather}
	   Y_i(\mathbf{z})= (\beta_0 + \epsilon_i)\cdot\frac{\deg_i}{\overline \deg} \cdot(
1 + \beta_1 z_i + r \frac{\sum_{j \in \mathcal{N}_i} z_j}{\deg_i}
), 
\\
    Y_i(\mathbf{z})=\beta_0+\beta_1 z_i+r \left(\frac{\sum_{j \in \mathcal{N}(i)} z_j}{\operatorname{deg}_i}\right)^2 + \epsilon_i.
\end{gather}
Here, $\epsilon_i \sim N(0,\sigma_e^2)$ are i.i.d. normal noises. We set the parameters as $(\beta_0,\beta_1,r,\sigma_e) = (1,1,1,0.1)$.

% \textbf{Regression model.} Throughout this section, we use a 3-layer GNN as the regression model, which is intentionally to be a lightweight one. The details of model architecture is reported in Appendix~\ref{app:exp_setting}. 
\textbf{Regression model.} Throughout this section, we intentionally use a lightweight 3-layer GNN as the regression model. The details of the model architecture are reported in Appendix~\ref{app:exp_setting}.
% BEGIN MOVED FROM APPENDIX: B1_GNN_ARCHITECTURE_AND_TRAINING
We choose the Chebyshev Spectral Graph Convolution (ChebConv)~\cite{defferrard2016convolutional} as our regression model, implemented through the Deep Graph Library~\cite{wang2019dgl}. Our GNN model consists of three ChebConv layers, with input dimension, output dimension, and filter size $(2, 16, 2), (16, 16, 1), (16, 1, 1)$, respectively. 

We train the GNN with Adam~\cite{kingma2014adam} to minimize MSE, using only treatment $z_i$ and degree $\deg_i$ as initial node features. In most settings, the training loss approaches the irreducible noise variance $\sigma_e^2$, indicating adequate optimization of the regression model.
% END MOVED FROM APPENDIX: B1_GNN_ARCHITECTURE_AND_TRAINING

\textbf{Merging scheme.} We use two schemes with $(c_1,\dots,c_5)=(2\%,5\%,10\%,25\%,50\%)$. The first fixes the final stage at $c_5=0.5$ and merges $\{c_{6-t},\dots,c_5\}$ for $t=1,\dots,5$, measuring the benefit of incorporating earlier ramp-up data. The second combines each $c_t$ with an initial stage that treats exactly one unit or cluster, approximating a zero-treatment baseline. This compares the current stage with an early stage. It is motivated by Theorem~\ref{theorem3}: under linear interference, merging the two extreme stages is already effective, while adding intermediate stages may increase bias.

\textbf{Randomization scheme.} We compare unit-level and cluster-level complete randomization under staggered rollout; Appendix~\ref{app:exp_setting} describes cluster construction. Appendix~\ref{app:exp_supp} also compares Bernoulli and complete randomization under both staggered rollout and temporally independent experiments, with similar performance across these alternatives.

\subsection{Results and Analysis}

\begin{figure}
	\centering
	\includegraphics[width=\columnwidth]{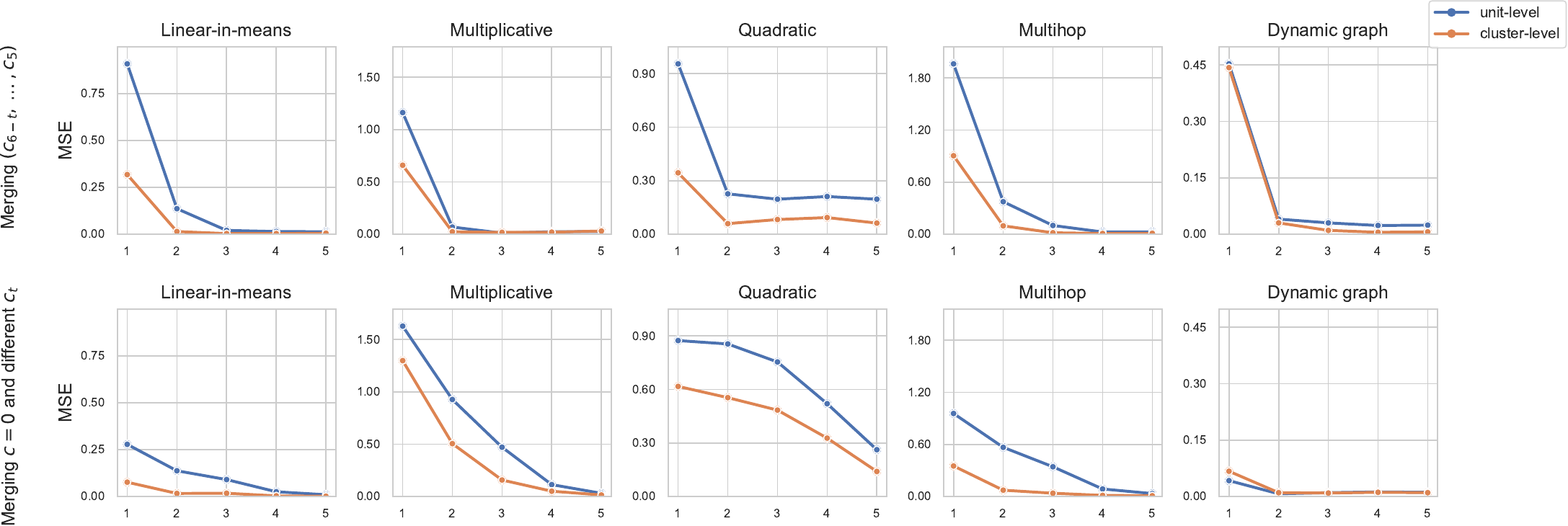}
	\caption{MSE of the GNN-based estimator under unit-level and cluster-level randomization. The top row merges stages $(c_{6-t},\dots,c_5)$; the bottom row combines each $c_t$ with an initial stage treating one unit or cluster ($c\approx0$). Both rows use the same vertical scale within each interference setting.}
	\label{fig:total_results}
\end{figure}

% We present the main results on MSE in Figure~\ref{fig:total_results}, and the numerical results of multihop interference setting in Table~\ref{table:multihop} for illustration. We leave detailed results in Appendix~\ref{app:exp_results} and additional experiments in Appendix~\ref{app:exp_supp}.
% We present the main MSE results in Figure~\ref{fig:total_results} and illustrate the multihop interference setting in Table~\ref{table:multihop}. Due to space limit, \textbf{detailed results and additional experiments} (including different randomization schemes, interference patterns, estimators without regression components, and network topology) are deferred to Appendices~\ref{app:exp_results} and~\ref{app:exp_supp}, respectively.
We report MSE in Figure~\ref{fig:total_results} and detailed 2-hop results in Table~\ref{table:multihop}. Appendices~\ref{app:exp_results} and~\ref{app:exp_supp} provide full numerical results and comparisons across randomization schemes, interference patterns, estimators without regression components, and network topologies.

\begin{table*}
    \centering
    \small
    \caption{Performance under \textbf{2-hop} interference for the two merging schemes.}
    \label{table:multihop}
    \begin{tabular}{l rrr rrr}
    \toprule
    & \multicolumn{3}{c}{Unit-level} & \multicolumn{3}{c}{Cluster-level} \\
    \cmidrule(lr){2-4} \cmidrule(lr){5-7}
    Stage & Bias & Std & MSE & Bias & Std & MSE \\
    \midrule
    \multicolumn{7}{l}{\textit{(a) Merging $t$ stages: $(c_{6-t},\dots,c_5)$}} \\
    $t=1$ & -1.401 & 0.034 & 1.963 & -0.950 & 0.045 & 0.904 \\
    $t=2$ & -0.611 & 0.024 & 0.374 & -0.303 & 0.064 & 0.096 \\
    $t=3$ & -0.316 & 0.013 & 0.100 & -0.117 & 0.046 & 0.016 \\
    $t=4$ & -0.159 & 0.015 & 0.025 & -0.061 & 0.064 & 0.008 \\
    $t=5$ & -0.160 & 0.024 & 0.026 & -0.074 & 0.070 & 0.010 \\
    \midrule
    \multicolumn{7}{l}{\textit{(b) Merging the initial stage ($c\approx0$) with $c_t$}} \\
    $c_t=0.02$ & -0.977 & 0.032 & 0.956 & -0.537 & 0.251 & 0.351 \\
    $c_t=0.05$ & -0.751 & 0.043 & 0.566 & -0.217 & 0.159 & 0.072 \\
    $c_t=0.10$ & -0.580 & 0.078 & 0.342 & -0.129 & 0.141 & 0.037 \\
    $c_t=0.25$ & -0.293 & 0.020 & 0.086 & -0.094 & 0.061 & 0.012 \\
    $c_t=0.50$ & -0.183 & 0.012 & 0.033 & -0.083 & 0.037 & 0.008 \\
    \bottomrule
    \end{tabular}
\end{table*}
Bias and standard deviation (Std) are estimated from 100 Monte Carlo repetitions. Std summarizes variability of the GATE estimate. The one-stage experiment at $c=0.5$ serves as the baseline, shown at $t=1$ in the top row of Figure~\ref{fig:total_results} and in Table~\ref{table:multihop}(a).

% We summarize the main conclusions as follows: 
% (i) \textbf{the benefits of merging two steps are intrinsic}, while the benefits of merging more intermediate steps (besides the lowest and largest treatment proportions) are often marginal; 
% (ii) \textbf{there exists a prominent synergistic effect between cluster-level randomization and our merging methodology}, as predicted in Section~\ref{ch4}, which we attribute to their joint role in increasing the variation of network exposure in the training data;
% (iii) \textbf{bias is indeed the dominant factor for the GNN-based estimator in most cases, and thus bias reduction leads to a significant improvement in estimator MSE}, e.g., see the first row of the left part of Table~\ref{table:multihop} and the tables in Appendix~\ref{app:exp_results}; 
% (iv) merging two steps with a larger treatment proportion gap induces more substantial bias reduction, e.g., see the right part of Table~\ref{table:multihop}; 
% (v) the benefits of merging are sourced from variation of proportions, and merging repeated experiments would be meaningless, e.g., see Appendix~\ref{app:exp_supp}.
The results support five conclusions:
(i) \textbf{two-step merging already provides substantial benefits};
(ii) \textbf{merging and cluster-level randomization act synergistically} by jointly enriching exposure variation, as anticipated in Section~\ref{ch4};
(iii) \textbf{bias dominates GNN estimation error in most settings}, so reducing it substantially improves MSE (Table~\ref{table:multihop}(a) and Appendix~\ref{app:exp_results});
(iv) \textbf{intermediate stages can help under complex interference}, including nonlinear models and dynamic graphs (Tables~\ref{table:quadratic_combined}--\ref{table:dynamic_combined});
(v) repeating experiments at a fixed proportion provides little bias reduction (Table~\ref{table:repeated}).

Table~\ref{table:multihop}(a) also shows that the gains need not be monotone. Under unit-level randomization, MSE falls from $1.963$ at $t=1$ to $0.374$ at $t=2$ and $0.025$ at $t=4$, while the fifth stage leaves it at $0.026$. Cluster-level randomization reaches an MSE of $0.008$ at $t=4$ and $0.010$ at $t=5$. Thus, additional stages can help, but their value depends on the exposure information they contribute. These patterns support the interpretation in Sections~\ref{ch3} and~\ref{ch4}; the GNN evidence complements the exact linear-regression analysis without extending its guarantees to arbitrary regressors.

% Overall, based on our comprehensive experiments, our practical recommendation is as follows: when training a regression-based estimator for GATE, one should use cluster-level randomization and merge at least two slices of experimental data with different treatment proportions.

% BEGIN MOVED FROM APPENDIX: B4_RANDOMIZATION_COMPARISON
\paragraph{Randomization schemes.} Independent Bernoulli randomization is common in practice because of its simplicity. Complete randomization induces only weak correlations among unit treatments and yields similar performance in our simulations. We also compare temporally independent experiments with staggered rollout, forming a $2\times2$ comparison reported in Tables~\ref{table:basic_bernoulli} and~\ref{table:basic_complete}.
% END MOVED FROM APPENDIX: B4_RANDOMIZATION_COMPARISON

% BEGIN MOVED FROM APPENDIX: B4_REPEATED_EXPERIMENTS
\paragraph{Merging repeated experiments.} To distinguish the effect of exposure variation from that of training-set size, we merge repeated experiments at a fixed treatment proportion.

Table~\ref{table:repeated} reports results under 2-hop interference when all repeated experiments treat 50\% of units. Repetition reduces standard deviation relative to one stage, but does not reduce bias or MSE. Increasing training-set size alone therefore does not reproduce the benefits of merging stages with \textbf{different} treatment proportions, consistent with Theorems~\ref{theorem2} and~\ref{theorem3}.

\begin{table}
    \centering
    \small
    \caption{Performance under 2-hop interference with \textbf{merged repeated experiments over $t$ steps}.}
    \label{table:repeated}
    \begin{tabular}{l ccc c ccc}
    \toprule
    & \multicolumn{3}{c}{{Unit}} & & \multicolumn{3}{c}{{Cluster}} \\
    \cmidrule(lr){2-4} \cmidrule(lr){6-8}
    {Rounds} & Bias & Std & MSE & & Bias & Std & MSE \\
    \midrule
    \textbf{$t=1$} & -1.402 & 0.034 & 1.968 & & -0.954 & 0.049 & 0.912 \\
    \textbf{$t=2$} & -1.460 & 0.015 & 2.132 & & -0.981 & 0.028 & 0.963 \\
    \textbf{$t=3$} & -1.480 & 0.012 & 2.189 & & -0.992 & 0.025 & 0.984 \\
    \textbf{$t=4$} & -1.486 & 0.012 & 2.208 & & -0.993 & 0.025 & 0.987 \\
    \textbf{$t=5$} & -1.488 & 0.012 & 2.214 & & -0.994 & 0.026 & 0.989 \\
    \bottomrule
    \end{tabular}
\end{table}
% END MOVED FROM APPENDIX: B4_REPEATED_EXPERIMENTS

Together, the experiments support combining cluster-level randomization with at least two stages having distinct treatment proportions when training a regression-based GATE estimator.

\section{Conclusion}

Ramp-up provides a statistical resource for GATE estimation in A/B tests under network interference. Our closed-form analysis shows that, under regular general linear interference, bias dominates the error of the linear regression estimator. Merging stages with different treatment proportions reduces this bias by increasing exposure variation. This perspective also explains the synergy with cluster-level randomization observed in our simulations. Across the studied interference settings, the benefits extend to lightweight GNN estimators, providing a practical way to improve counterfactual prediction using existing ramp-up data.

The conclusions concern static GATE and do not establish unbiasedness under arbitrary interference. Under a fixed traffic cap, merging can still leave substantial bias, as Equation~\eqref{eq:traffic_cap_bias} illustrates. Our closed-form analysis holds potential-outcome parameters fixed across stages and excludes cross-round spillovers; the nonlinear and dynamic-graph results provide simulation evidence rather than a general guarantee. In deployment, the available stages must also satisfy operational risk and traffic constraints. Stage selection should therefore consider exposure diversity as well as sample size, with the earliest and latest usable stages providing a natural starting point. Extending the analysis to carryover effects and adaptive rollout decisions remains an important direction.

\pagebreak
\bibliography{ref}
\bibliographystyle{plain}

%%%%%%%%%%%%%%%%%%%%%%%%%%%%%%%%%%%%%%%%%%%%%%%%%%%%%%%%%%%%

\newpage
\appendix

\section{Additional Discussion on Theoretical Results}\label{app:discussion}

\subsection{Discussion on Assumption~\ref{assumption1}}\label{app:discuss_assump1}

In this subsection, we revisit Assumption~\ref{assumption1} and discuss its implications and practicality.

Scenarios where interference is excessively strong are relatively rare. Moreover, to assess the relative order of variance compared to bias in such cases, we will inevitably encounter the order of $\sum_{(i,j,k,l) \in \mathcal{I}} B_{ij} B_{kl} / \sum_{i,j} B_{ij}$ for various index sets $\mathcal{I}$ as outlined in Theorem~\ref{theorem1}. This complexity makes it challenging to derive succinct conclusions without specifying the interference matrix $B$.

For a specific unit $i$, the interference on its outcome is represented by $ \sum_{j} B_{ij} z_j$, and the treatment effect is $\beta_1$. The first part of Assumption~\ref{assumption1} restricts the interference effect received by unit $i$ to be at most of the constant order, ensuring that interference does not overshadow the outcome. Additionally, this assumption allows for some units to be minimally affected by interference. Finally, through basic algebra, one can verify that $\sum_{j} |B_{ij}|$ is equal to $1$ in Example~\ref{example1} and bounded above by $\sum_{m=1}^M r_m$ in Example~\ref{example2}.

The second part of Assumption~\ref{assumption1} restricts the order of total interference effect to be regular. While we assume that $\sum_{i,j} B_{ij} = \Theta(n)$, the bias reduction results discussed in the subsequent subsections also apply when $\sum_{i,j} B_{ij} = o(n)$, which indicates very weak interference. We do not incorporate this case here because the bias introduced by interference diminishes as $n$ increases, given that $\sum_{i,j} B_{ij}/n = o(1)$. As illustrated in the expression of bias, it is unnecessary to use specific techniques designed to address interference, in this case. Therefore, we focus on the meaningful and challenging scenario, where $\sum_{i,j} B_{ij} = \Theta(n)$. For instance, in Examples~\ref{example1} and~\ref{example2}, the order of $\sum_{ij} B_{ij}$ are both $\Theta(n)$.

The graph-structural interpretation of this condition is given in Section~\ref{ch3}. This assumption is particularly reasonable for typical social networks, where, even with over a billion nodes, it is unlikely for any single unit to have thousands of friends while maintaining significant influence over each one of them.

\subsection{Discussion on Theorem~\ref{theorem2}}\label{app:discuss_theorem2}

Lowering $c_1$ and increasing $c_2$ brings us closer to these target mean outcomes. The traffic-capped bias expression is now given in Equation~\eqref{eq:traffic_cap_bias}. Moreover, as demonstrated in the next subsection, merging multistep experimental data does not lead to further bias reduction compared to the two-step case. Therefore, we choose to explore more powerful regression models, specifically GNNs, in our simulation study.

\subsection{Discussion on Theorem~\ref{theorem3}}\label{app:discuss_theorem3}

In machine learning practice, it is common to see that merging more related data enhances the performance of the predictive models. However, through basic algebra, one can verify that merging more experimental data does \textbf{not} guarantee further bias reduction. This raises a natural question: Given the completed experiments with treatment proportions $c_1<c_2<\dots<c_T$, which new experiment with proportion $x$ will further reduce bias? We address this question through the following corollary. 
\begin{corollary}\label{corollary2}
    To further reduce the bias, $x$ should satisfy: 
    \begin{equation}   
    \left(T^2C_1 - (T+1)C_1^2 +TC_2\right) x^2 - \left((T-1)C_1^2 +2TC_1C_2 + T(T+1)C_2\right)x + C_1^3 - C_1^2C_2 > 0,
    \end{equation}
    here $C_1 = \sum_{t=1}^T   c_t, \  
    C_2 = \sum_{t=1}^T   c_t^2$    
\end{corollary}

For this quadratic function with respect to $x$, we notice the following two facts about the coefficient of the quadratic term and the constant term:
\begin{equation}
    \begin{aligned}
         T^2C_1 - (T+1)C_1^2 +TC_2 &= \left(TC_1(T-C_1)\right) + \left(TC_2 - C_1^2\right) >0 \\ 
        C_1^3 - C_1^2C_2 &= C_1^2 (C_1-C_2) >0.
    \end{aligned}
\end{equation}
We then draw several conclusions about the quadratic function in Corollary~\ref{corollary2}. First, when $c_1 > 0$, merging data with $x = 0$—representing data under global control—can always reduce bias. In practice, this type of data is almost always accessible on online platforms. Second, intuitively, a beneficial proportion $x$ should either be lower than $c_1$ or higher than $c_T$, rather than lying within the middle section of $[c_1, c_T]$. This intuition is only for illustration and not rigorous, since $x$ around $c_1$ or $c_T$ may also bring slight bias reduction, which depends on the specific values of $(c_t)_{t=1}^T  $. Considering the first point discussed above, an effective way to further reduce bias is to conduct a larger-scale experiment, i.e., with $x > c_T$.

Additionally, we can apply Corollary \ref{corollary2} in reverse: given the treatment proportions $\{c_i\}_{i=1}^T/\{c_t\}$, removing data with the treatment proportion $c_t$ from the merged dataset may further reduce bias. Repeating this procedure suggests that the optimal solution may involve merging only the datasets with proportions $c_1$ and $c_T$. This finding highlights the unique importance of the $T=2$ case, which we will further validate in the simulation study. Nonetheless, with nonlinear interference terms, merging multistep experimental data can offer advantages, as will be discussed in Section~\ref{ch5}.

While it is theoretically possible to solve an optimization problem to select the treatment proportions $(c_t)_{t=1}^T  $ for maximizing bias reduction, we do not recommend this approach. First, the optimization problem is dependent on the specific potential outcome model and the regression model used. Although we provide a detailed analysis of the simple regression estimator, it is not typically used in practice. Second, when selecting ramp-up steps in real-world experiments, the experimenter must consider factors such as risk and budget in advance, while such an optimization problem does not capture these practical considerations.

Finally, we discuss our proposed intuition of increasing exposure variation, as introduced in Section~\ref{ch4}. The corresponding numerical illustration is reported in Section~\ref{ch4}.

\subsection{Discussion on Temporal Interference}

The focus of this paper is network interference, namely the cross-sectional component. Although temporal interference may arise across multiple experiments, it is not our primary concern, and we abstract it away in our analysis. In principle, our analytical framework can be extended to accommodate both temporal interference and dynamic graph structures; however, doing so would alter the target estimand---the static GATE---and complicate the main narrative of this paper. In short, we study the regime in which cross-sectional interference is more substantial than temporal interference and therefore deserves primary attention.

\section{Detailed Experiments Results}\label{app:exp}

\subsection{Details of Basic Setting}\label{app:exp_setting}

\paragraph{Data.} The main network topology is a Facebook social network used in many related works, available at \url{https://networkrepository.com/socfb-Stanford3.php}.

\paragraph{GNN.} The model architecture and optimization objective are described in Section~\ref{ch5}.

 Other miscellaneous parameters are given below:

\begin{enumerate}
    % \item \textcolor{red}{what this means? what's the sample size for policy evaluation?} 
    \item Resolution of clustering: 10
    \item Number of clusters: 192
    \item Activation function: ReLU
    \item Number of training epochs: 400
    \item Optimizer: Adam with learning rate 0.004 and momentum coefficient (betas) 0.9,0.999.
\end{enumerate}

\paragraph{Potential outcome models.}

Beyond the results of the standard linear-in-means model (Example~\ref{example1}), we examine three non-linear interference patterns to demonstrate the generality of our theoretical insights, including \textbf{multiplicative}, \textbf{quadratic}, and \textbf{square-root interference}. 
\begin{gather}
	   Y_i(\mathbf{z})= (\beta_0 + \epsilon_i)\cdot\frac{\deg_i}{\overline \deg} \cdot(
1 + \beta_1 z_i + r \frac{\sum_{j \in \mathcal{N}_i} z_j}{\deg_i}
) 
\\
    Y_i(\mathbf{z})=\beta_0+\beta_1 z_i+r \left(\frac{\sum_{j \in \mathcal{N}(i)} z_j}{\operatorname{deg}_i}\right)^2 + \epsilon_i
\\ 
Y_i(\mathbf{z})=\beta_0+\beta_1 z_i+r 
    \sqrt{\left(\frac{\sum_{j \in \mathcal{N}(i)} z_j}{\operatorname{deg}_i}\right)} + \epsilon_i,
\end{gather}
Here, $\epsilon_i \sim N(0,\sigma_e^2)$ are i.i.d. normal noises. We set the parameters as $(\beta_0,\beta_1,r,\sigma_e) = (1,1,1,0.1)$.
We defer the introduction of our dynamic graph structure to Appendix~\ref{app:dynamic_intro}.

\subsection{Detailed Results}\label{app:exp_results}

We present the main numerical results below. Table~\ref{table:linear_interference_combined} shows that the theoretical conclusions derived in Section~\ref{ch3} and Appendix~\ref{app:discussion} continue to hold for the GNN regressor. In particular, merging the first and last ramp-up stages ($c_1,c_T$) is already highly effective, and can even outperform merging intermediate stages. Under more complex interference patterns, however, incorporating additional intermediate stages can be substantially more beneficial, much like two points suffice to determine a line, whereas a cubic curve requires four.

\begin{table*}
    \centering
    \scriptsize
    \renewcommand{\arraystretch}{0.8}
    \caption{\textbf{Linear-in-means} model: (left) merge $t$ stages; (right) merge $c=0$ and $c_t$.}
    \label{table:linear_interference_combined}
    \resizebox{\textwidth}{!}{
    \begin{tabular}{l ccc ccc c l ccc ccc}
    \toprule
    & \multicolumn{3}{c}{Unit} & \multicolumn{3}{c}{Cluster} & & & \multicolumn{3}{c}{Unit} & \multicolumn{3}{c}{Cluster} \\
    \cmidrule(lr){2-4} \cmidrule(lr){5-7} \cmidrule(lr){10-12} \cmidrule(lr){13-15}
    \textbf{Rounds} & Bias & Std & MSE & Bias & Std & MSE & & \textbf{Proportion} & Bias & Std & MSE & Bias & Std & MSE \\
    \midrule
    \textbf{$t=1$} & -0.951 & 0.071 & 0.909 & -0.561 & 0.059 & 0.318 & & \textbf{$c_t=0.02$} & -0.526 & 0.034 & 0.278 & -0.216 & 0.173 & 0.076 \\
    \textbf{$t=2$} & -0.369 & 0.021 & 0.136 & -0.112 & 0.037 & 0.014 & & \textbf{$c_t=0.05$} & -0.369 & 0.025 & 0.137 & -0.032 & 0.121 & 0.016 \\
    \textbf{$t=3$} & -0.142 & 0.014 & 0.020 & -0.034 & 0.044 & 0.003 & & \textbf{$c_t=0.10$} & -0.299 & 0.028 & 0.090 & -0.010 & 0.130 & 0.017 \\
    \textbf{$t=4$} & -0.116 & 0.023 & 0.014 & -0.030 & 0.048 & 0.003 & & \textbf{$c_t=0.25$} & -0.155 & 0.020 & 0.025 & -0.037 & 0.045 & 0.003 \\
    \textbf{$t=5$} & -0.110 & 0.020 & 0.013 & -0.042 & 0.054 & 0.005 & & \textbf{$c_t=0.50$} & -0.092 & 0.015 & 0.009 & -0.015 & 0.033 & 0.001 \\
    \bottomrule
    \end{tabular}
    }
\end{table*}

\begin{table*}
    \centering
    \scriptsize
    \renewcommand{\arraystretch}{0.8}
    \caption{\textbf{Multiplicative} model: (left) merge $t$ stages; (right) merge $c=0$ and $c_t$.}
    \label{table:multiplicative_combined}
    \resizebox{\textwidth}{!}{
    \begin{tabular}{l ccc ccc c l ccc ccc}
    \toprule
    & \multicolumn{3}{c}{Unit} & \multicolumn{3}{c}{Cluster} & & & \multicolumn{3}{c}{Unit} & \multicolumn{3}{c}{Cluster} \\
    \cmidrule(lr){2-4} \cmidrule(lr){5-7} \cmidrule(lr){10-12} \cmidrule(lr){13-15}
    \textbf{Rounds} & Bias & Std & MSE & Bias & Std & MSE & & \textbf{Proportion} & Bias & Std & MSE & Bias & Std & MSE \\
    \midrule
    \textbf{$t=1$} & -1.078 & 0.039 & 1.163 & -0.811 & 0.046 & 0.659 & & \textbf{$c_t=0.02$} & -1.275 & 0.055 & 1.629 & -1.103 & 0.288 & 1.299 \\
    \textbf{$t=2$} & -0.257 & 0.037 & 0.068 & -0.122 & 0.089 & 0.023 & & \textbf{$c_t=0.05$} & -0.959 & 0.082 & 0.927 & -0.666 & 0.249 & 0.505 \\
    \textbf{$t=3$} & -0.061 & 0.077 & 0.010 & -0.058 & 0.115 & 0.017 & & \textbf{$c_t=0.10$} & -0.681 & 0.082 & 0.471 & -0.340 & 0.205 & 0.157 \\
    \textbf{$t=4$} & -0.110 & 0.106 & 0.023 & -0.003 & 0.137 & 0.019 & & \textbf{$c_t=0.25$} & -0.321 & 0.098 & 0.113 & -0.193 & 0.111 & 0.050 \\
    \textbf{$t=5$} & -0.120 & 0.104 & 0.025 & -0.118 & 0.125 & 0.030 & & \textbf{$c_t=0.50$} & -0.164 & 0.058 & 0.030 & -0.101 & 0.063 & 0.014 \\
    \bottomrule
    \end{tabular}
    }
\end{table*}

\begin{table*}
    \centering
    \scriptsize
    \renewcommand{\arraystretch}{0.8}
    \caption{\textbf{Quadratic} interference: (left) merge $t$ stages; (right) merge $c=0$ and $c_t$.}
    \label{table:quadratic_combined}
    \resizebox{\textwidth}{!}{
    \begin{tabular}{l ccc ccc c l ccc ccc}
    \toprule
    & \multicolumn{3}{c}{Unit} & \multicolumn{3}{c}{Cluster} & & & \multicolumn{3}{c}{Unit} & \multicolumn{3}{c}{Cluster} \\
    \cmidrule(lr){2-4} \cmidrule(lr){5-7} \cmidrule(lr){10-12} \cmidrule(lr){13-15}
    \textbf{Rounds} & Bias & Std & MSE & Bias & Std & MSE & & \textbf{Proportion} & Bias & Std & MSE & Bias & Std & MSE \\
    \midrule
    \textbf{$t=1$} & -0.976 & 0.044 & 0.955 & -0.583 & 0.063 & 0.344 & & \textbf{$c_t=0.02$} & -0.932 & 0.067 & 0.874 & -0.776 & 0.120 & 0.617 \\
    \textbf{$t=2$} & -0.475 & 0.020 & 0.226 & -0.240 & 0.040 & 0.059 & & \textbf{$c_t=0.05$} & -0.924 & 0.032 & 0.855 & -0.739 & 0.084 & 0.554 \\
    \textbf{$t=3$} & -0.443 & 0.010 & 0.196 & -0.284 & 0.038 & 0.082 & & \textbf{$c_t=0.10$} & -0.868 & 0.021 & 0.754 & -0.692 & 0.074 & 0.484 \\
    \textbf{$t=4$} & -0.460 & 0.015 & 0.211 & -0.301 & 0.044 & 0.093 & & \textbf{$c_t=0.25$} & -0.721 & 0.023 & 0.520 & -0.569 & 0.058 & 0.327 \\
    \textbf{$t=5$} & -0.442 & 0.032 & 0.196 & -0.241 & 0.064 & 0.062 & & \textbf{$c_t=0.50$} & -0.512 & 0.010 & 0.262 & -0.373 & 0.035 & 0.140 \\
    \bottomrule
    \end{tabular}
    }
\end{table*}

\begin{table*}
    \centering
    \scriptsize
    \renewcommand{\arraystretch}{0.8}
    \caption{\textbf{Square-root} interference: (left) merge $t$ stages; (right) merge $c=0$ and $c_t$.}
    \label{table:sqrt_combined}
    \resizebox{\textwidth}{!}{
    \begin{tabular}{l ccc ccc c l ccc ccc}
    \toprule
    & \multicolumn{3}{c}{Unit} & \multicolumn{3}{c}{Cluster} & & & \multicolumn{3}{c}{Unit} & \multicolumn{3}{c}{Cluster} \\
    \cmidrule(lr){2-4} \cmidrule(lr){5-7} \cmidrule(lr){10-12} \cmidrule(lr){13-15}
    \textbf{Rounds} & Bias & Std & MSE & Bias & Std & MSE & & \textbf{Proportion} & Bias & Std & MSE & Bias & Std & MSE \\
    \midrule
    \textbf{$t=1$} & -0.905 & 0.042 & 0.821 & -0.623 & 0.051 & 0.391 & & \textbf{$c_t=0.02$} & -0.324 & 0.027 & 0.106 & 0.218 & 0.304 & 0.140 \\
    \textbf{$t=2$} & -0.387 & 0.014 & 0.150 & -0.180 & 0.046 & 0.035 & & \textbf{$c_t=0.05$} & -0.090 & 0.109 & 0.020 & 0.696 & 0.241 & 0.543 \\
    \textbf{$t=3$} & -0.185 & 0.012 & 0.034 & -0.063 & 0.052 & 0.007 & & \textbf{$c_t=0.10$} & 0.519 & 0.050 & 0.272 & 0.895 & 0.145 & 0.822 \\
    \textbf{$t=4$} & -0.086 & 0.010 & 0.008 & 0.021 & 0.057 & 0.004 & & \textbf{$c_t=0.25$} & 0.528 & 0.030 & 0.280 & 0.579 & 0.091 & 0.344 \\
    \textbf{$t=5$} & -0.006 & 0.012 & 0.000 & 0.030 & 0.081 & 0.007 & & \textbf{$c_t=0.50$} & 0.218 & 0.008 & 0.048 & 0.250 & 0.057 & 0.066 \\
    \bottomrule
    \end{tabular}
    }
\end{table*}

\begin{table*}
    \centering
    \scriptsize
    \renewcommand{\arraystretch}{0.8}
    \caption{\textbf{Dynamic graph structure}: (left) merge $t$ stages; (right) merge $c=0$ and $c_t$.}
    \label{table:dynamic_combined}
    \resizebox{\textwidth}{!}{
    \begin{tabular}{l ccc ccc c l ccc ccc}
    \toprule
    & \multicolumn{3}{c}{Unit} & \multicolumn{3}{c}{Cluster} & & & \multicolumn{3}{c}{Unit} & \multicolumn{3}{c}{Cluster} \\
    \cmidrule(lr){2-4} \cmidrule(lr){5-7} \cmidrule(lr){10-12} \cmidrule(lr){13-15}
    \textbf{Rounds} & Bias & Std & MSE & Bias & Std & MSE & & \textbf{Proportion} & Bias & Std & MSE & Bias & Std & MSE \\
    \midrule
    \textbf{$t=1$} & -0.666 & 0.095 & 0.453 & -0.665 & 0.025 & 0.443 & & \textbf{$c_t=0.02$} & -0.196 & 0.061 & 0.042 & -0.220 & 0.135 & 0.067 \\
    \textbf{$t=2$} & -0.150 & 0.132 & 0.040 & -0.161 & 0.061 & 0.030 & & \textbf{$c_t=0.05$} & -0.069 & 0.048 & 0.007 & -0.086 & 0.055 & 0.010 \\
    \textbf{$t=3$} & -0.106 & 0.137 & 0.030 & -0.094 & 0.028 & 0.010 & & \textbf{$c_t=0.10$} & -0.092 & 0.037 & 0.010 & -0.084 & 0.041 & 0.009 \\
    \textbf{$t=4$} & -0.060 & 0.141 & 0.023 & -0.058 & 0.044 & 0.005 & & \textbf{$c_t=0.25$} & -0.107 & 0.021 & 0.012 & -0.100 & 0.025 & 0.011 \\
    \textbf{$t=5$} & -0.045 & 0.149 & 0.024 & -0.034 & 0.070 & 0.006 & & \textbf{$c_t=0.50$} & -0.099 & 0.031 & 0.011 & -0.098 & 0.028 & 0.010 \\
    \bottomrule
    \end{tabular}
    }
\end{table*}

\subsection{Dynamic Graph Structure}\label{app:dynamic_intro}

We introduce our setting of a dynamic graph structure, which is novel in the literature but common in practice. For instance, if each experiment is conducted over a period of one week, the edges of the graph can change within that week, even in social networks. New friendships may form, and existing ones may be dissolved. The evolution of network structure is well-documented in network science, with models such as the Barabási-Albert Model~\cite{barabasi1999emergence} and the Forest Fire Model~\cite{leskovec2005graphs}. Although this aspect is still underexplored in the context of network experiments, it is valuable to consider this practical factor in sequential multiple experiments, such as in the ramp-up process.

\begin{remark}
    Here, for simplicity, we consider the exogenous network evolution that is independent of treatment assignment. This is because the treatments typically considered by online platforms, such as new product features, generally do not have a significant impact on social relationships. However, there is existing literature in economics that examines cases where treatment does endogenously influence network formation, usually when the treatment itself is more consequential. 
\end{remark}

For simplicity, we keep the node set unchanged throughout the ramp-up process and focus on modeling the addition of new edges over time. Inspired by classic preferential attachment, we generate edges for nodes with degrees in the top 50\%. For each such node $i$, we connect it to node $j$ with the following probability:
\begin{equation}
 p_{ij} = \frac{\deg_j}{\sum_k \deg_k},
\end{equation}
In each evolution, we repeat this sampling procedure 5 times with replacement for each node with degrees in the top 50\%. If a unit $i$ is sampled when generating edges for itself, we discard this sampling to avoid self-loops. Additionally, if the sampled unit $j$ is already connected to unit $i$ at the current step, we also discard this sampling to avoid creating multiple edges.

Given the ramp-up steps $(c_1,c_2,\dots,c_5)=(2\%, 5\%, 10\%, 25\%, 50\%)$, we take the initial network topology $\mathcal{G}$ as baseline and repeat such evolution 5 times to generate evolved graphs with the same node set. The number of edges:
\begin{equation}
 |\mathcal{E}_1|,\dots, |\mathcal{E}_5| = 596414, 624465, 652391, 680313, 708274,
\end{equation}
where $\mathcal{E}_t$ is the edge set at the $t$-th step. 

Moreover, we go back to use the classic linear-in-means model and examine the influence of dynamic graph structure. The main parameter is set as $(\beta_0,\beta_1,r,\sigma_e) = (1,1,1,0.1)$, as in the previous settings. Even if the graph structure evolves, the GATE remains $\beta_1+r=2$. We present the result in Table~\ref{table:dynamic_combined}.

The impact of evolutionary changes is quite complex, with classic models typically only characterizing the limiting behavior of preferential attachment. Nonetheless, our methodology consistently enhances the GNN model's performance. During the evolution process, we preserve the clustering of the original topology $\mathcal{G}$ to implement the staggered rollout design\footnote{The time complexity of the Louvain algorithm supports weekly updates for clustering on large online platforms. However, when clustering changes, monotonic treatment assignment may no longer be guaranteed under cluster-level randomization.}, although this clustering may become less effective as edges are continuously added. As a result, our suggested practice of merging data with global control and the highest treatment proportion faces challenges due to the increasing discrepancy between $\mathcal{G}_t$ and $\mathcal{G}$ as step $t$ progresses. This provides an intuitive explanation for why the monotonic improvement observed with increasing $c_t$, as shown in our basic experiments, no longer persists in Table~\ref{table:dynamic_combined}. This reveals a potential limitation of this merging strategy in practice.

\begin{remark}
    The rationale for generating edges solely for high-degree nodes stems from the distribution of treatment exposure, as detailed in the main paper. Low-degree nodes typically have exposure values concentrated at the boundaries of the interval $(0,1)$, which significantly contributes to the variation in exposures. To preserve this distinctive structure, we intentionally avoid generating new edges for all nodes. However, we do allow low-degree nodes to receive new edges initiated by high-degree nodes.
\end{remark}

\subsection{Supplementary Experiments}\label{app:exp_supp}

The randomization comparison is discussed in Section~\ref{ch5}; the complete numerical results are reported below.

\begin{table*}
    \centering
    \scriptsize
    \renewcommand{\arraystretch}{0.8}
    \caption{Performance under \textbf{independent Bernoulli randomization}: (Left) \textbf{temporal independence}, (Right) \textbf{staggered rollout}.}
    \label{table:basic_bernoulli}
    \resizebox{\textwidth}{!}{
    \begin{tabular}{l ccc ccc c l ccc ccc}
    \toprule
    & \multicolumn{3}{c}{Unit} & \multicolumn{3}{c}{Cluster} & & & \multicolumn{3}{c}{Unit} & \multicolumn{3}{c}{Cluster} \\
    \cmidrule(lr){2-4} \cmidrule(lr){5-7} \cmidrule(lr){10-12} \cmidrule(lr){13-15}
    \textbf{Rounds} & Bias & Std & MSE & Bias & Std & MSE & & \textbf{Rounds} & Bias & Std & MSE & Bias & Std & MSE \\
    \midrule
    \textbf{$t=1$} & -0.945 & 0.071 & 0.899 & -0.562 & 0.058 & 0.320 & & \textbf{$t=1$} & -0.944 & 0.072 & 0.896 & -0.560 & 0.058 & 0.317 \\
    \textbf{$t=2$} & -0.365 & 0.023 & 0.133 & -0.118 & 0.046 & 0.016 & & \textbf{$t=2$} & -0.365 & 0.022 & 0.134 & -0.117 & 0.047 & 0.016 \\
    \textbf{$t=3$} & -0.142 & 0.015 & 0.020 & -0.038 & 0.041 & 0.003 & & \textbf{$t=3$} & -0.142 & 0.015 & 0.020 & -0.036 & 0.044 & 0.003 \\
    \textbf{$t=4$} & -0.115 & 0.024 & 0.014 & -0.032 & 0.050 & 0.004 & & \textbf{$t=4$} & -0.116 & 0.023 & 0.014 & -0.033 & 0.051 & 0.004 \\
    \textbf{$t=5$} & -0.110 & 0.019 & 0.012 & -0.042 & 0.056 & 0.005 & & \textbf{$t=5$} & -0.110 & 0.019 & 0.012 & -0.042 & 0.056 & 0.005 \\
    \bottomrule
    \end{tabular}
    }
\end{table*}

\begin{table*}
    \centering
    \scriptsize
    \renewcommand{\arraystretch}{0.8}
    \caption{Performance under \textbf{complete randomization}: (Left) \textbf{temporal independence}, (Right) \textbf{staggered rollout}.}
    \label{table:basic_complete}
    \resizebox{\textwidth}{!}{
    \begin{tabular}{l ccc ccc c l ccc ccc}
    \toprule
    & \multicolumn{3}{c}{Unit} & \multicolumn{3}{c}{Cluster} & & & \multicolumn{3}{c}{Unit} & \multicolumn{3}{c}{Cluster} \\
    \cmidrule(lr){2-4} \cmidrule(lr){5-7} \cmidrule(lr){10-12} \cmidrule(lr){13-15}
    \textbf{Rounds} & Bias & Std & MSE & Bias & Std & MSE & & \textbf{Rounds} & Bias & Std & MSE & Bias & Std & MSE \\
    \midrule
    \textbf{$t=1$} & -0.953 & 0.071 & 0.914 & -0.560 & 0.059 & 0.317 & & \textbf{$t=1$} & -0.951 & 0.071 & 0.909 & -0.561 & 0.059 & 0.318 \\
    \textbf{$t=2$} & -0.369 & 0.021 & 0.136 & -0.114 & 0.037 & 0.014 & & \textbf{$t=2$} & -0.369 & 0.021 & 0.136 & -0.112 & 0.037 & 0.014 \\
    \textbf{$t=3$} & -0.141 & 0.015 & 0.020 & -0.033 & 0.045 & 0.003 & & \textbf{$t=3$} & -0.142 & 0.014 & 0.020 & -0.034 & 0.044 & 0.003 \\
    \textbf{$t=4$} & -0.115 & 0.024 & 0.014 & -0.031 & 0.047 & 0.003 & & \textbf{$t=4$} & -0.116 & 0.023 & 0.014 & -0.030 & 0.048 & 0.003 \\
    \textbf{$t=5$} & -0.110 & 0.020 & 0.013 & -0.042 & 0.054 & 0.005 & & \textbf{$t=5$} & -0.110 & 0.020 & 0.013 & -0.042 & 0.054 & 0.005 \\
    \bottomrule
    \end{tabular}
    }
\end{table*}

\paragraph{Merging repeated experiments.} In Theorem~\ref{theorem2} and \ref{theorem3}, we conclude that the potential bias reduction of merging experiments with the same treatment proportion should be very marginal, i.e., at most of order $O(1/n)$, and we examine this for the linear regression estimator here.

The empirical comparison is reported in Table~\ref{table:repeated} in the main text.

\paragraph{Linear regression estimator} We then supplement the statistical performance of the linear regression estimator under the basic setting, as shown in Figure~\ref{fig:combined}. As in the main paper, we fix the randomization scheme as complete randomization with staggered rollout. We consider treatment proportions $(c_1,c_2,\dots,c_5)=(2\%, 5\%, 10\%, 25\%, 50\%)$ in the ramp-up process and train on merged data from $t=1,2,3,4,5$ steps, corresponding to treatment proportions $(c_{6-t},\dots, c_5)$. The detailed results are reported in Table~\ref{table:lr}. By basic algebra, these numerical results validate Theorem~\ref{theorem2}.

\begin{table}
    \centering
    \scriptsize
    \renewcommand{\arraystretch}{0.8}
    \caption{Performance with \textbf{linear regression estimator} when merging $t$ steps.}
    \label{table:lr}
    \begin{tabular}{l ccc c ccc}
    \toprule
    & \multicolumn{3}{c}{{Unit}} & & \multicolumn{3}{c}{{Cluster}} \\
    \cmidrule(lr){2-4} \cmidrule(lr){6-8}
    {Rounds} & Bias & Std & MSE & & Bias & Std & MSE \\
    \midrule
    \textbf{$t=1$} & -1.000 & 0.004 & 1.000 & & -0.595 & 0.010 & 0.354 \\
    \textbf{$t=2$} & -0.933 & 0.003 & 0.871 & & -0.554 & 0.019 & 0.307 \\
    \textbf{$t=3$} & -0.866 & 0.003 & 0.750 & & -0.518 & 0.027 & 0.269 \\
    \textbf{$t=4$} & -0.824 & 0.003 & 0.680 & & -0.493 & 0.028 & 0.244 \\
    \textbf{$t=5$} & -0.792 & 0.003 & 0.628 & & -0.471 & 0.031 & 0.223 \\
    \bottomrule
    \end{tabular}
\end{table}

\paragraph{Lagrange interpolation estimator}
We next supplement the performance of the Lagrange interpolation estimator proposed in~\cite{cortez2022agnostic}. Unlike our regression-based approach, this is a system-level estimator that is agnostic to the graph microstructure: it approximates the one-dimensional mean outcome curve $\bar Y(p)$ as a function of the treatment proportion $p$ under unit-level independent Bernoulli randomization. Its polynomial basis effectively assumes a low-order interaction structure of the form
\begin{equation}
    \sum_{\ell=1}^\beta w_l \left(\frac{\sum_{j \in \mathcal{N}_i} \tilde{c}_{i j} z_j}{\sum_{j \in \mathcal{N}_i} \tilde{c}_{i j}}\right)^{\ell},
\end{equation}
where $\tilde{c}_{ij}$ characterizes the influence of unit $j$ on unit $i$, and $w_\ell$ is the coefficient of the $\ell$-th order term.

We note that Lagrange interpolation estimator is a \textbf{graph-agnostic} estimator that summarizes the experimental data at each timestep into a single value---mean outcomes. Thus, it heavily relies on the exact knowledge of $\beta$ and incurs misspecification risk for both potential outcome model and \textbf{number of timesteps} (note that the order of polynomial in Lagrange interpolation is often exactly $T-1$ due to random error). We report its performance on two settings: multiplicative interference in Table~\ref{table:lagrange_multi} and square-root interference in Table~\ref{table:lagrange_sqrt}.

In the multiplicative setting, we observe that Lagrange interpolation performs very well for the case of $t=2$, while it rapidly deteriorates with the increase of $t$, which is counterintuitive but can be explained by our summary: \textbf{it heavily relies on the exact knowledge on $\beta$}, which is usually inaccessible in practice. Moreover, although the theoretical derivation of both our paper and ~\cite{cortez2022agnostic} are based on unit-level randomization, the functional form of Lagrange interpolation estimator there substantially relies on the randomization, while our idea is generalizable (through the lens of increasing exposure variation), i.e., improving the data for learning representation. Therefore, we observe a general synergistic effect with cluster-level randomization and our merging methodology, while Lagrange interpolation estimator performs even much worse on experimental data generated with cluster-level randomization.

In the square-root setting (the performance of GNN-based estimator with our methodology is provided in Table~\ref{table:sqrt_combined}), which corresponds a scenario of diminishing marginal utility, our methodology benefits a lot from merging more steps (left part of Table~\ref{table:sqrt_combined}), while 

We summarize that our GNN-based approach learns a flexible node-level representation from all units directly, and can naturally accommodate richer microstructure, nodal covariates, and more complex interaction patterns.

\begin{table}
    \centering
    \scriptsize
    \renewcommand{\arraystretch}{0.8}
    \caption{\textbf{Lagrange interpolation} under \textbf{multiplicative} interference with $t$ stages.}
    \label{table:lagrange_multi}
    \begin{tabular}{l ccc c ccc}
    \toprule
    & \multicolumn{3}{c}{{Unit}} & & \multicolumn{3}{c}{{Cluster}} \\
    \cmidrule(lr){2-4} \cmidrule(lr){6-8}
    {Rounds} & Bias & Std & MSE & & Bias & Std & MSE \\
    \midrule
    \textbf{$t=1$} & N/A & N/A & N/A & & N/A & N/A & N/A \\
    \textbf{$t=2$} & -0.002 & 0.036 & 0.001 & & 0.010 & 0.237 & 0.056 \\
    \textbf{$t=3$} & -0.005 & 0.074 & 0.005 & & 0.045 & 0.479 & 0.231 \\
    \textbf{$t=4$} & 0.013 & 0.561 & 0.315 & & 0.740 & 3.501 & 12.806 \\
    \textbf{$t=5$} & 0.045 & 8.437 & 71.192 & & 4.802 & 49.852 & 2508.301 \\
    \bottomrule
    \end{tabular}
\end{table}

\begin{table}
    \centering
    \scriptsize
    \renewcommand{\arraystretch}{0.8}
    \caption{\textbf{Lagrange interpolation} under \textbf{square-root} interference with $t$ stages.}
    \label{table:lagrange_sqrt}
    \begin{tabular}{l ccc c ccc}
    \toprule
    & \multicolumn{3}{c}{{Unit}} & & \multicolumn{3}{c}{{Cluster}} \\
    \cmidrule(lr){2-4} \cmidrule(lr){6-8}
    {Rounds} & Bias & Std & MSE & & Bias & Std & MSE \\
    \midrule
    \textbf{$t=1$} & N/A & N/A & N/A & & N/A & N/A & N/A \\
    \textbf{$t=2$} & -0.131 & 0.015 & 0.017 & & -0.079 & 0.331 & 0.116 \\
    \textbf{$t=3$} & -0.387 & 0.032 & 0.150 & & -0.282 & 0.685 & 0.549 \\
    \textbf{$t=4$} & 1.276 & 0.366 & 1.761 & & 1.787 & 5.708 & 35.769 \\
    \textbf{$t=5$} & -18.021 & 7.106 & 375.254 & & -8.481 & 88.365 & 7880.220 \\
    \bottomrule
    \end{tabular}
\end{table}

\paragraph{Network-aware estimators without regression component.} We next provide the performance of two network-aware estimators without regression component for comparison, including MII~\cite{chen2026journey} and Hájek estimator~\cite{ugander2023randomized}, in Table~\ref{table:stat_estimators}. We empirically find that naively averaging the estimates acquired on different timesteps $t$ brings no improvement, so we only report the performance with treatment proportion $t=50\%$. This table is provided only for comparison.

\begin{table}
    \centering
    \scriptsize
    \renewcommand{\arraystretch}{0.8}
    \caption{Performance of MII and Hájek estimators under 2-hop interference.}
    \label{table:stat_estimators}
    \begin{tabular}{l ccc c ccc}
    \toprule
    & \multicolumn{3}{c}{{Unit}} & & \multicolumn{3}{c}{{Cluster}} \\
    \cmidrule(lr){2-4} \cmidrule(lr){6-8}
    {Estimator} & Bias & Std & MSE & & Bias & Std & MSE \\
    \midrule
    {MII}   & N/A & N/A & N/A & & -0.276 & 0.010 & 0.077 \\
    {Hájek} & -0.471 & 0.043 & 0.223 & & -0.263 & 0.039 & 0.071 \\
    \bottomrule
    \end{tabular}
\end{table}

\paragraph{Another network topology.} For completeness, we supplement the performance of GNN-based estimator on MIT8\footnote{The network topology is available at \url{https://networkrepository.com/socfb-MIT8.php}.}~\cite{nrfbstanf}. We examine the case of 2-hop interference, since it is concerned with the concrete topology. We report the results in Table~\ref{table:mit8_2hop}, and the results demonstrate that our methodology remains effective.

\begin{table*}
    \centering
    \scriptsize
    \renewcommand{\arraystretch}{0.8}
    \caption{\textbf{2-hop} interference on \textbf{MIT8}: (left) merge $t$ stages; (right) merge $c=0$ and $c_t$.}
    \label{table:mit8_2hop}
    \resizebox{\textwidth}{!}{
    \begin{tabular}{l ccc ccc c l ccc ccc}
    \toprule
    & \multicolumn{3}{c}{Unit} & \multicolumn{3}{c}{Cluster} & & & \multicolumn{3}{c}{Unit} & \multicolumn{3}{c}{Cluster} \\
    \cmidrule(lr){2-4} \cmidrule(lr){5-7} \cmidrule(lr){10-12} \cmidrule(lr){13-15}
    \textbf{Rounds} & Bias & Std & MSE & Bias & Std & MSE & & \textbf{Proportion} & Bias & Std & MSE & Bias & Std & MSE \\
    \midrule
    \textbf{$t=1$} & -1.439 & 0.023 & 2.072 & -1.022 & 0.025 & 1.044 & & \textbf{$c_t=0.02$} & -0.880 & 0.052 & 0.776 & -0.467 & 0.256 & 0.284 \\
    \textbf{$t=2$} & -0.729 & 0.018 & 0.531 & -0.411 & 0.052 & 0.172 & & \textbf{$c_t=0.05$} & -0.633 & 0.052 & 0.403 & -0.246 & 0.119 & 0.075 \\
    \textbf{$t=3$} & -0.351 & 0.018 & 0.123 & -0.170 & 0.033 & 0.030 & & \textbf{$c_t=0.10$} & -0.545 & 0.032 & 0.298 & -0.192 & 0.055 & 0.040 \\
    \textbf{$t=4$} & -0.208 & 0.019 & 0.044 & -0.101 & 0.031 & 0.011 & & \textbf{$c_t=0.25$} & -0.292 & 0.031 & 0.086 & -0.088 & 0.051 & 0.010 \\
    \textbf{$t=5$} & -0.177 & 0.025 & 0.032 & -0.086 & 0.039 & 0.009 & & \textbf{$c_t=0.50$} & -0.184 & 0.019 & 0.034 & -0.091 & 0.022 & 0.009 \\
    \bottomrule
    \end{tabular}
    }
\end{table*}

\section{Proofs}\label{app:proof}

\subsection{Proof of Theorem \ref{theorem1}}\label{proof_of_theo1}

\textbf{Proof}. First, our simple linear regression estimator gives:
\begin{equation}
    \hat \tau = (\hat\theta_1 + \hat\theta_0) - \hat\theta_0 = \hat\theta_1,
\end{equation}
and 
\begin{equation}
    \hat\theta =  (X(\mathbf{z})^\top  X(\mathbf{z}))^{-1}X(z)^\top  Y,
\end{equation}

For the simplicity in the derivation, we write $X=X(\mathbf{z})$ in the following. We recap that $X(\mathbf{z})=\begin{pmatrix}
    \mathbf{1} &\mathbf{z}
\end{pmatrix}$. Since we consider complete randomization, there are $n$ units and we assign $d$ treatments exactly. We have:
\begin{equation}
    X^\top  X = 
    \begin{pmatrix}
        n & \mathbf{z}^\top  \mathbf{1} \\
        \mathbf{z}^\top  \mathbf{1} & \|\mathbf{z}\|_2^2
    \end{pmatrix}
    = 
    \begin{pmatrix}
        n & d \\
        d & d    
    \end{pmatrix},
\end{equation}
We remark that, for precise analysis on the bias and variance of linear regression estimator, it is critical to remove the randomness in $\det\left(X(\mathbf{z})^TX(\mathbf{z})\right)$.

Naturally,
\begin{equation}
    (X^\top  X)^{-1}X^\top   
    = 
    \frac{1}{d(n-d)}
    \begin{pmatrix}
        d(\mathbf{1}-\mathbf{z})^\top   \\ (n\mathbf{z}-d\mathbf{1})^\top      
    \end{pmatrix},    
\end{equation}

Notice that $Y$ can be written as 
\begin{equation}
    Y = \begin{pmatrix}
        X & B\mathbf{z}
    \end{pmatrix}  
    \beta,
\end{equation}
Here, $\beta = \begin{pmatrix}
    \beta_0 & \beta_1&1
\end{pmatrix}$. The shape of $X$ is $n \times 2$, $B\mathbf{z}$ is $n \times 1$, and $\beta$ is $3 \times 1$.

The expression of $\hat \theta$ is given by 
\begin{equation}
    \begin{aligned}
        \hat\theta &= (X^\top  X)^{-1} X^\top   Y \\ 
        &= 
        \begin{pmatrix}
            I_2  & (X^\top  X)^{-1} X^\top   B\mathbf{z}
        \end{pmatrix}
        \beta + (X^\top  X)^{-1} X^\top  \mathbf{\epsilon} \\            
        &= 
        \begin{pmatrix}
            I_2  & \frac{1}{d(n-d)} 
            \begin{pmatrix}
                d(\mathbf{1}-\mathbf{z})^\top   B\mathbf{z} \\ (n\mathbf{z}-d\mathbf{1})^\top   B\mathbf{z}    
            \end{pmatrix}            
        \end{pmatrix}
        + 
        (X^\top  X)^{-1} X^\top  \mathbf{\epsilon}.
    \end{aligned}
\end{equation}
Since we're only concerned with $\hat{\theta}_1$, we directly take the second term of $\hat{\theta}$:
\begin{equation}
    \hat\theta_1 = \beta_1 +  \frac{1}{d(n-d)} (n\mathbf{z}-d\mathbf{1})^\top  B\mathbf{z} + \frac{1}{d(n-d)} (n\mathbf{z}-d\mathbf{1})^\top  \epsilon,
\end{equation}

Notice that $\mathbf{z} \perp \mathbf{\epsilon}$. The \textbf{bias} is given by
\begin{equation}
    \mathbb{E}[\hat\theta_1] - (\beta_1+ \frac{1}{n}\operatorname{trace}(\mathbf{1}\mathbf{1}^\top B)) = \frac{1}{nc(1-c)}\mathbb{E}[(\mathbf{z} - c\mathbf{1})^\top   B\mathbf{z}] - \frac{1}{n}\operatorname{trace}(\mathbf{1}\mathbf{1}^\top B),
\end{equation}

Next, we evaluate the term $\mathbb{E}[(\mathbf{z} - c\mathbf{1})^\top   B\mathbf{z}]$, which is given by:
\begin{equation}\label{eq:zbz}
    \begin{aligned}
        \mathbb{E}[(\mathbf{z} - c\mathbf{1})^\top   B\mathbf{z}] 
        &=
        \mathbb{E}\left[\operatorname{trace}\left( (\mathbf{z}-c\mathbf{1})^\top   B\mathbf{z} \right)\right] 
        \\ 
        &= 
        \mathbb{E}\left[\operatorname{trace}\left( \mathbf{z}(\mathbf{z}-c\mathbf{1})^\top   B \right)\right]
        \\
        &= 
        \operatorname{trace}\left( \mathbb{E}\left[\mathbf{z}(\mathbf{z}-c\mathbf{1})^\top  \right] B \right)
        \\
        &= 
        \operatorname{trace}\left( (c \frac{d-1}{n-1} - c^2)\mathbf{1}\mathbf{1}^\top  B \right) 
        \\ 
        &= 
        -\frac{c(1-c)}{n-1} \operatorname{trace}(\mathbf{1}\mathbf{1}^\top  B).
    \end{aligned}   
\end{equation}

Plugging it into the expression of bias, we get:
\begin{equation}
    \text{Bias}(\hat \tau_{(1)}) = -\frac{\sum_{i,j} B_{ij}}{n}\left(1+\frac{1}{n-1}\right) =-\frac{\sum_{i,j} B_{ij}}{n-1}.
\end{equation}

Subsequently, we derive the expression of variance. Noting that $\mathbf{z} \perp \mathbf{\epsilon}$ and $\mathbb{E}[\mathbf{\epsilon}]=\mathbf{0}$, we have:
\begin{equation}
    \operatorname{Cov}[ (n\mathbf{z}-d\mathbf{1})^\top  B\mathbf{z}, (n\mathbf{z}-d\mathbf{1})^\top  \mathbf{\epsilon}]= 0,
\end{equation}

Thus, we can omit all the covariance terms in the expression of variance, yielding:
\begin{equation}
    \operatorname{Var}(\hat \tau) = 
        (\frac{1}{nc(1-c)})^2 \operatorname{Var}\left((\mathbf{z}-c\mathbf{1})^\top  B\mathbf{z}\right)  
        + 
        \frac{1}{\left(nc(1-c)\right)^2}
        \mathbb{E}[(\mathbf{z}-c\mathbf{1})^\top  (\mathbf{z}-c\mathbf{1})] \sigma^2_e, 
\end{equation}

First, we consider the second term above, which represents the variance induced by $\mathbf{\epsilon}$:
\begin{equation}
    \begin{aligned}
        \operatorname{Var}_e
        &=\frac{1}{(d(n-d))^2} \mathbb{E}[(n\mathbf{z}-d\mathbf{1})^\top  (n\mathbf{z}-d\mathbf{1})] \sigma^2_e 
        \\
        &= 
        \left(\frac{n}{d(n-d)}\right)^2
        \mathbb{E}[(\mathbf{z}-\frac{d}{n}\mathbf{1})^\top  (\mathbf{z}-\frac{d}{n}\mathbf{1})] \sigma^2_e
        \\
        &=
        \left(\frac{n\sigma_e}{d(n-d)}\right)^2
        \mathbb{E}[d - \frac{2d}{n}\times d +\frac{d^2}{n}] 
        \\
        &=
        \left(\frac{n\sigma_e}{d(n-d)}\right)^2
        (d-\frac{d^2}{n})
        \\
        &= \frac{\sigma_e^2}{nc(1-c)}.
        \end{aligned}
\end{equation}

Next, we discuss the first term: 
\begin{equation}
    \left(\frac{1}{nc(1-c)}\right)^2 \operatorname{Var}\left((\mathbf{z}-c\mathbf{1})^\top  B\mathbf{z}\right),
\end{equation}
We use the expansion $\operatorname{Var}(X) = \mathbb{E}[X^2] - E[X]^2$, and the square of expectation term has been simplified in equation (\ref{eq:zbz}). Thus, our focus is the expectation of square term, which can be expanded with brute-force: 
\begin{equation}
    \left(\frac{1 }{c(1-c)n}\right)^2\sum_{i,j,k,l}B_{ij}B_{kl}\mathbb{E}[(z_i-c)(z_k-c)z_j z_l],
\end{equation}

As mentioned in the main paper, our technique involves classifying the index set $I \subset [n]^4$ that restricts the quadruples $(i,j,k,l)$ and calculating $\mathbb{E}[(z_i-c)(z_k-c)z_j z_l]$ respectively. Since $B_{ii} = 0$ for all $i \in [n]$, we do not need to consider the cases where $i = j$ or $k = l$. We present the result as follows:
\begin{enumerate}
    \item $i=k,j=l,i\neq k$, 
    \begin{equation}
        \begin{aligned}
            \mathbb{E}[(z_i-c)(z_k-c)z_j z_l] 
            &= \mathbb{E}[(z_i-c)^2z_j^2 ] 
            \\
            &= \mathbb{E}[(z_i-2cz_i+c^2)z_j ]
            \\
            &= (1-2c)\mathbb{E}[z_iz_j ] + c^2\mathbb{E}[z_j ]
            \\ 
            &= (1-2c)c \frac{d-1}{n-1} + c^3      
            \\
            &= 
            c^2(1-c) + O\left(\frac{1}{n}\right).
            \end{aligned}   
    \end{equation}

    \item $i=l,k=j,i\neq k$ 
    \begin{equation}
        \begin{aligned}
            \mathbb{E}[(z_i-c)(z_k-c)z_j z_l] 
            &= \mathbb{E}[(z_i-c)(z_k-c)z_iz_k ] 
            \\
            &= \mathbb{E}[(1-c)^2z_iz_k]
            \\
            &= (1-c)^2c \frac{d-1}{n-1}  
            \\
            &= 
            (1-c)^2c^2 + O\left(\frac{1}{n}\right).
            \end{aligned}
    \end{equation}

    \item Only $i=k$ holds, with $i$, $j$, and $l$ being distinct. 
    \begin{equation}
        \begin{aligned}
            \mathbb{E}[(z_i-c)(z_k-c)z_j z_l] 
            &= 
            \mathbb{E}[(z_i-c)^2z_kz_l] 
            \\
            &= 
            (1-2c)\mathbb{E}[z_iz_kz_l]  + c^2 \mathbb{E}[z_kz_l]
            \\
            &= 
            (1-2c)c \frac{(d-1)(d-2)}{(n-1)(n-2)}  + c^3 \frac{d-1}{n-1}  
            \\
            &= 
            c^3(1-c) + O\left(\frac{1}{n}\right).
            \end{aligned}
    \end{equation}    

    \item Only $j=l$ holds, with $i$, $j$, and $k$ being distinct. 
    \begin{equation}\label{eq:coef_second_term_of_var}
        \begin{aligned}
            \mathbb{E}[(z_i-c)(z_k-c)z_j z_l]  
            &= \mathbb{E}[(z_i-c)(z_k-c)z_j] 
            \\ 
            &= \frac{nd(d-n)+2d^2+2n^2-4nd}{n^2(n-1)(n-2)}
            \\
            &= \frac{c(d-n) + 2c^2 + 2 - 4c}{(n-1)(n-2)}
            \\
            &= 
            \frac{c(c-1)}{n}+O(\frac{1}{n^2}).
            \end{aligned}
    \end{equation}
    
    \item $i,j,k,l$ are all distinct.
    \begin{equation}
        \begin{aligned}
            &\phantom{=} \ \ \mathbb{E}[(z_i-c)(z_k-c)z_j z_l] 
            \\
            &= \mathbb{E}[(z_iz_k-c(z_i+z_k)+c^2)z_jz_l ] 
            \\
            &= 
            \frac{d(d-1)(d-2)(d-3)}{n(n-1)(n-2)(n-3)} - 2c \frac{d(d-1)(d-2)}{n(n-1)(n-2)} + c^2 \frac{d(d-1)}{n(n-1)}
            \\
            &= 
            \frac{d(d-1)}{n(n-1)} 
            \left(
            \frac{(d-2)(d-3)}{(n-2)(n-3)} + \frac{d^2}{n^2} - 2 \frac{d(d-2)}{n(n-2)}
            \right)
            \\
            &= 
            \frac{d(d-1)}{n(n-1)} 
            \left(
            \frac{(d^2-5d+6)n^2 + d^2(n^2-5d+6)-2dn(dn-2n-3d+6)}{n^2(n-2)(n-3)}
            \right)
            \\
            &= 
            \frac{d(d-1)}{n(n-1)} 
            \left(
            \frac{(d^2-5d+6)n^2 + d^2(n^2-5d+6)-2dn(dn-2n-3d+6)}{n^2(n-2)(n-3)}
            \right)
            \\
            &=
            \frac{c(d-1)}{(n-1)} 
            \left(
            \frac{c(d-n)+6+6c^2-12c}{(n-2)(n-3)}
            \right)    
            \\
            &= 
            \frac{c^3(c-1)}{n} +O(\frac{1}{n^2}).
            \end{aligned}
    \end{equation}
\end{enumerate}

Combine these five cases and corresponding index sets, we get the expression of variance as presented in Theorem 1. For the convenience of proof of Corollary \ref{corollary1} later, we also annotate the order of each term at last line. 
\qed

% app1.2
\subsection{Proof of Corollary \ref{corollary1}}\label{proof_of_coro1}
\textbf{Proof}. Since Assumption~\ref{assumption1} restricts 
\begin{equation}
    \sum_{i, j} B_{i j}=\Theta(n),
\end{equation}
We get the order of bias instantly:
\begin{equation}
    \begin{aligned}
        \text{Bias}(\hat \tau_{(1)})
        = -\frac{\sum_{i,j} B_{ij}}{n-1} 
        = \Theta(1).
    \end{aligned}
\end{equation}

Next, we discuss the order of variance. It concerns bounding the order of the summations over five index sets mentioned in Appendix~\ref{proof_of_theo1}. To begin with, we notice that the variance is non-negative and the variance induced by $\epsilon$ is:
\begin{equation}
    \frac{\sigma_e^2}{ n c(1-c)} = \Theta(\frac{1}{n}),
\end{equation}
Thus, we can first claim that 
\begin{equation}
    \operatorname{Var}(\hat \tau) = \Omega(\frac{1}{n}),
\end{equation}

We then prove the first term in the expression of variance has a order of $O(1/n)$. Specifically, we bound those five summations respectively. According to the Assumption~\ref{assumption1}, we get a naive bound that will be used repeatedly:
\begin{equation}
    |B_{ij}| \leq \sum_{j}|B_{ij}| \leq b. 
\end{equation}

\begin{enumerate}
    \item $i=k,j=l,i\neq k$. 
    \begin{equation}
        \sum_{ij} B_{ij}^2 \leq \sum_{i} b\sum_{j} |B_{ij}| \leq b^2 n,
    \end{equation}
    Thus, 
    \begin{equation}
        \sum_{ij} B_{ij}^2 = O(n).
    \end{equation}

    \item $i=l,k=j,i\neq k$ 
    \begin{equation}
        |\sum_{i, j} B_{i j} B_{j i}| \leq \sum_{i, j} |B_{i j} B_{j i}| \leq b \sum_{i,j}B_{ij} \leq b^2 n,
    \end{equation}
    Thus, 
    \begin{equation}
        \sum_{ij} B_{ij}B_{ji} = O(n).
    \end{equation}    

    \item Only $i=k$ holds, with $i$, $j$, and $l$ being distinct. 
    \begin{equation}
        |\sum_{i, j, l} B_{i j} B_{i l}| 
        \leq 
        \sum_{i, j, l} |B_{i j} B_{i l}| 
        = 
        \sum_{i, j} |B_{i j}| \sum_{l} |B_{i l}| 
        \leq 
        b \sum_{i, j} |B_{i j}|
        \leq 
        b^2n,
    \end{equation}  
    Thus, 
    \begin{equation}
        \sum_{i, j, l} B_{i j} B_{i l} = O(n).
    \end{equation}

    \item Only $j=l$ holds, with $i$, $j$, and $k$ being distinct.  
    \begin{equation}
        |\sum_{i, j, k} B_{i j} B_{k j}| 
        \leq 
        \sum_{i, j, k} |B_{i j} B_{k j}| 
        \leq 
        b\sum_{i, j, k} |B_{i j}|
        \leq 
        b^2 n^2,
    \end{equation}
    Thus, 
    \begin{equation}
        \sum_{i, j, k} B_{i j} B_{k j} = O(n^2).
    \end{equation}

    \item $i,j,k,l$ are all distinct.
    \begin{equation}
        \sum_{i,j,k,l} B_{ij} B_{kl} = (\sum_{ij} B_{ij})^2 = \Theta(n^2).     
    \end{equation}

\end{enumerate}

Combine the orders of five summations above and the order of corresponding $\mathbb{E}[(z_i-c)(z_k-c)z_j z_l]$, we get:
\begin{equation}
    \operatorname{Var}(\hat{\tau}) = \Theta(\frac{1}{n}),
\end{equation}

Finally, we comment that the bound on $|\sum_{i, j, k} B_{i j} B_{k j}|$ is rough, while we do not apply the third part of Assumption~\ref{assumption1}, i.e. restricting the order $\sum_{i, j, k} B_{i j} B_{k j}=\sum_j\left(\sum_i B_{i j}\right)^2$ directly. 

Such a rough bound is still acceptable because of the equation (\ref{eq:coef_second_term_of_var}), i.e.
\begin{equation}
    \mathbb{E}[(z_i-c)(z_k-c)z_j z_l] = \Theta(\frac{1}{n}),        
\end{equation}
For the situation of merging $T\geq 2$ experimental data, if $\frac{\sum_{t=1}^T   c_t}{T} \neq \mathbb{E}[z_i]$, we would have:
\begin{equation}
    \mathbb{E}[(z_i- \frac{\sum_{t=1}^T   c_t}{T})(z_k - \frac{\sum_{t=1}^T   c_t}{T})z_j z_l] = \Theta(1),
\end{equation}
This explains why we need a further assumption to control the order of variance there.
\qed

% app1.3
\subsection{Proof of Theorem \ref{theorem2}}\label{proof_of_theo2}
\textbf{Proof}. In the situation of $T=2$, the sample size increases from $n$ to $2n$, and the number of treated units changes from $d$ to $d_1 + d_2$. Correspondingly, the treatment proportions in two experiments are $c_1$ and $c_2$. In linear regression, this additionally leads to the following two changes:
\begin{equation}
    \begin{aligned}
        B\rightarrow 
        \begin{pmatrix}
            B & 0\\ 0& B    
        \end{pmatrix}
        \qquad 
        \mathbf{z} \rightarrow 
        \begin{pmatrix}
            \mathbf{z}_1 \\ \mathbf{z}_2
        \end{pmatrix}.
    \end{aligned}
\end{equation}

We make the changes above and get 
\begin{equation}\label{eq:bias_2step}
    \begin{aligned}
        &\phantom{=}\ \ \text{Bias}(\hat \tau_{(2)})/(\sum_{i,j}B_{ij}) + 1\\
        &= 
        \frac{2n}{(d_1+d_2)(2n-(d_1+d_2))} \operatorname{trace}\left( \mathbb{E}\left[\mathbf{z}(\mathbf{z}-\frac{d_1+d_2}{2n}\mathbf{1})^\top  \right] 
        \begin{pmatrix}
            B & 0\\ 0& B
        \end{pmatrix}  \right)
        \\
        &= 
        \frac{2n}{(d_1+d_2)(2n-(d_1+d_2))} \operatorname{trace}\left( 
        \left(\mathbb{E}\left[\mathbf{z}\mathbf{z^\top  }\right]
        -\frac{d_1+d_2}{2n}\mathbb{E}\left[\mathbf{z}\mathbf{1}^\top \right] \right)
        \begin{pmatrix}
            B & 0\\ 0& B
        \end{pmatrix} \right)
        \\ 
        &=
        \frac{2n}{(d_1+d_2)(2n-(d_1+d_2))} 
        \left(
        \begin{aligned}
        &\operatorname{trace}
        \left((\mathbb{E}[\mathbf{z}_1\mathbf{z}_1^\top  ]+\mathbb{E}[\mathbf{z}_2\mathbf{z}_2^\top  ])B \right)
        \\
        &\quad -
        \frac{d_1+d_2}{2n}(\frac{d_1}{n}+\frac{d_2}{n})\operatorname{trace}\left(\mathbf{1}\mathbf{1}^\top  B \right)
        \end{aligned}
        \right)
        \\
        &=
        \frac{2n}{(d_1+d_2)(2n-(d_1+d_2))} 
        \left(
        \frac{d_1(d_1-1)+d_2(d_2-1)}{n(n-1)} - \frac{(d_1+d_2)^2}{2n^2}
        \right) \operatorname{trace}\left(\mathbf{1}\mathbf{1}^\top  B \right)
        \\
        &= 
        \frac{2n^2}{(d_1+d_2)(2n-(d_1+d_2))} 
        \frac{(n-1)(d_1-d_2)^2+2(d_1^2+d_2^2)-2n(d_1+d_2)}{2n^2(n-1)}
        \\
        &=
        \frac{(n-1)(d_1-d_2)^2+2(d_1^2+d_2^2)-2n(d_1+d_2)}{(n-1)(d_1+d_2)(2n-(d_1+d_2))}
        \\
        &= 
        \frac{(n-1)\left(c_1-c_2\right)^2+2\left(c_1^2+c_2^2\right)-2\left(c_1+c_2\right)}{(n-1)\left(c_1+c_2\right)\left(2-\left(c_1+c_2\right)\right)}.
        \end{aligned}
\end{equation}

Thus, we get:
\begin{equation}
    \operatorname{Bias}(\hat \tau_{(2)}) =-\frac{\sum_{i j} B_{i j}}{n}\left(1-\frac{(n-1)\left(c_1-c_2\right)^2+2\left(c_1^2+c_2^2\right)-2\left(c_1+c_2\right)}{(n-1)\left(c_1+c_2\right)\left(2-\left(c_1+c_2\right)\right)}\right).
\end{equation}

% app1.4
\subsection{Proof of Theorem \ref{theorem3}}\label{proof_of_theo3}
\textbf{Proof}. As mentioned in the main paper, there is no additional challenge in deriving the result of $T>2$ case. Refer to the fourth equation in equation (\ref{eq:bias_2step}), we now have: 
\begin{equation}
    \begin{aligned}
        \text{Bias}\!(\hat \tau_{(T)})/( \sum_{i,j}B_{ij}) + 1
        &=
        \frac{Tn}{\sum_{t=1}^T   d_t(Tn-\sum_{t=1}^T   d_t)}
        \left(\frac{\sum_{t=1}^T   d_t(d_t-1)}{n(n-1)} - \frac{(\sum_{t=1}^T   d_t)^2}{Tn^2}\right)
        \\
        &= 
        \frac{T}{n(\sum_{t=1}^T   c_t) (T-\sum_{t=1}^T   c_t)}  \left(\frac{T\sum_{t=1}^T   c_t^2 - (\sum_{t=1}^T   c_t)^2}{T} + O\left(\frac{1}{n}\right)\right).
        \end{aligned}
\end{equation}
Thus, since $\sum_{i,j}B_{ij} = \Theta(n)$, we have:
\begin{equation}
    \begin{aligned}
        \text{Bias}(\hat \tau_{(T)}) &= 
        -
        \frac{\sum_{ij}B_{ij}}{n} 
        \left( 1 - \frac{T\sum_{t=1}^T   c_t^2 - (\sum_{t=1}^T   c_t)^2}{(\sum_{t=1}^T   c_t)(T - \sum_{t=1}^T   c_t)} \right) + O\left(\frac{1}{n}\right).
    \end{aligned}
\end{equation}

\subsection{Proof of Corollary \ref{corollary2}}\label{proof_of_coro2}

\textbf{Proof}. We continue to use the notation in the main paper, and further bias reduction corresponds to:
\begin{equation}
    \frac{kC_2 - C_1^2}{C_1(k - C_1)} < \frac{(k+1)(C_2+x^2) - (C_1+x)^2}{(C_1+x)(k+1 - C_1-x)},
\end{equation}

We expand the two sides of this inequality with brute-force, and get following condition:
\begin{equation}
    \begin{aligned}
        &\phantom{\geq}
        k(k+1)C_1C_2 + k(k+1)C_1x^2 - (k+1)C_1^2C_2 - (k+1)C_1^2 x^2 
        \\
        &- kC_1^3 - 2kC_1^2 x-kC_1x^2 +C_1^4 + 2C_1^3x +C_1^2 x^2 
        \\
        &> 
        k(k+1)C_1C_2 + k(k+1)C_2x - kC_1^2 C_2 -2k C_1C_2x 
        \\
        &- kC_2x^2 - (k+1)C_1^3 +C_1^4 +C_1^3x -(k+1)C_1^2x +C_1^3x + C_1^2x^2
        ,
        \end{aligned}
\end{equation}
which is equivalent to:
\begin{equation}
    \left(k^2C_1 - (k+1)C_1^2 +kC_2\right) x^2 - \left((k-1)C_1^2 +2kC_1C_2 + k(k+1)C_2\right)x + C_1^3 - C_1^2C_2 >0.
\end{equation}
\qed

\subsection{Lemma \ref{lemma1} for Variance Analysis with Proof}\label{proof_of_lemm1}

We construct the following lemma for the convenience of variance analysis:
\begin{lemma}\label{lemma1}
    Given Assumption~\ref{assumption1}, for a treatment vector $\mathbf{z}$ with treatment proportion $c\in(0,1)$ and any constant $w\in (0,1)$, we have:
    \begin{equation}
        \operatorname{Var}\left(  
          (\mathbf{z}-w\mathbf{1})^\top B\mathbf{z}\right) = O\left(\frac{1}{n}\right).
    \end{equation}
\end{lemma}

\textbf{Proof}. First, we will continue to apply the idea of expansion in the proof of variance part of Theorem 1. However, the proof in the Theorem 1 can't apply directly because the order of $\mathbb{E}[(z_i-c)(z_k-c)z_j z_l]$ increases from $\Theta(1/n)$ to $\Theta(1)$, when $w\neq c$. These two cases are:
\begin{enumerate}
    \item Only $j=l$ holds, with $i$, $j$, and $k$ being distinct.   
    \item $i,j,k,l$ are all distinct.
\end{enumerate}

For the first case, we resort to the third part of Assumption~\ref{assumption1}:
\begin{equation}
    \sum_{i, j, k} B_{i j} B_{k j}=\sum_j(\sum_i B_{i j})^2=O(n),
\end{equation}

For the second case, we will prove that the two coefficients of term $(\sum_{ij}B_{ij})^2$ in the expression of $\operatorname{Var}((\mathbf{z}-w \mathbf{1})^\top   B \mathbf{z})$ can cancel out. Namely, we will prove:
\begin{equation}
    \mathbb{E}[(z_i-w)(z_k-w)z_j z_l] - \left(\mathbb{E}[(\mathbf{z}-w \mathbf{1})^\top   B \mathbf{z}]/\operatorname{trace}(\mathbf{1}\mathbf{1}^\top  B)\right)^2 = O\left(\frac{1}{n}\right),
\end{equation}
for the case $i,j,k,l$ are not equal to each other.

Notice that, for $s=1,2,3$, we have:
\begin{equation}
    \frac{d-s}{n-s} = c + O\left(\frac{1}{n}\right),
\end{equation}
 
On one hand, 
\begin{equation}
    \begin{aligned}
        \mathbb{E}[(z_i-w)(z_k-w)z_j z_l] 
        &= 
        \mathbb{E}[(z_i-w)(z_k-w)\mid z_j=z_l=1] \left(c^2 + O\left(\frac{1}{n}\right)\right)
        \\
        &= 
        (c-w)^2c^2 + O\left(\frac{1}{n}\right).
    \end{aligned}    
\end{equation}

On the other hand, 
\begin{equation}
    \begin{aligned}
        \frac{\mathbb{E}[(\mathbf{z}-w \mathbf{1})^\top   B \mathbf{z}]}{\operatorname{trace}(\mathbf{1}\mathbf{1}^\top  B)}
        &=         
        \frac{\operatorname{trace}\left(\mathbb{E}\left[\mathbf{z}(\mathbf{z}-w \mathbf{1})^\top  \right] B\right)}{\operatorname{trace}(\mathbf{1}\mathbf{1}^\top  B)}
        \\
        &= \frac{\operatorname{trace}\left((\mathbb{E}[\mathbf{z}\mathbf{z}^\top  ]-cw \mathbf{1}\mathbf{1}^\top  )B\right)}{\operatorname{trace}(\mathbf{1}\mathbf{1}^\top  B)} 
        \\
        &= c(c-w) + O\left(\frac{1}{n}\right).
    \end{aligned}
\end{equation}

In the last equation, we again utilize $B_{ii}=0$ for all unit $i\in[n]$. Combine two equations above, we successfully control the order of coefficient of term $(\sum_{ij}B_{ij})^2$, in the expression of concerned variance $\operatorname{Var}((\mathbf{z}-w \mathbf{1})^\top   B \mathbf{z})$.

For other three cases of $\mathbb{E}[(z_i-c)(z_k-c)z_j z_l]$, the proof of Theorem 1 can be utilized directly here, because $\mathbb{E}[(z_i-c)(z_k-c)z_j z_l]=\Theta(1)$ there, and the order of this term can't exceed constant no matter what value $w$ takes. Therefore, we finish the proof.
\qed

\subsection{Proof of Theorem \ref{theorem4}}\label{proof_of_theo4}

\textbf{Proof}. We first present the proof for the case of \textbf{temporally independent} $T$ experiments is a direct extension of Lemma~\ref{lemma1}. In fact, with Lemma~\ref{lemma1}, we can conclude that 
\begin{equation}
    \operatorname{Var}\left(\sum_{t=1}^T  \left(\mathbf{z}_t-\frac{\sum_{k=1}^\top   c_k}{T} \mathbf{1}\right)^\top   B \mathbf{z}_t\right)=\sum_{t=1}^T   \operatorname{Var}\left(\left(\mathbf{z}_t-\frac{\sum_{k=1}^\top   c_k}{T} \mathbf{1}\right)^\top   B \mathbf{z}_t\right) = O\left(\frac{1}{n}\right),
\end{equation}

On the other hand, the order of variance introduced by $\epsilon$ is always $\Theta(1/n)$, and the concerned variance is the sum of these non-negative variance terms. Therefore, we can conclude that 
\begin{equation}
    \operatorname{Var}(\hat{\tau}) =\Theta\left(\frac{1}{n}\right),
\end{equation}

% \subsection{Proof of Theorem 5}\label{proof_of_theo5}

Next, we study the case of \textbf{staggered rollout} experiments. The complexity here is strong temporal covariance of treatments.  

Based on the Lemma~\ref{lemma1}, the remaining work is bounding the sum of covariance terms. We have:
\begin{equation}
    \begin{aligned}
        \operatorname{Var}\left(
          \sum_{t=1}^T  
          (\mathbf{z}_t-\frac{\sum_{k=1}^\top   c_k}{T}\mathbf{1})^\top  B\mathbf{z}_t
        \right)
        &= 
        \sum_{t=1}^T  
        \operatorname{Var}\left(  
          (\mathbf{z}_t-\frac{\sum_{k=1}^\top   c_k}{T}\mathbf{1})^\top  B\mathbf{z}_t
        \right) 
        \\
        + 
        2\sum_{t_1< t_2}\operatorname{Cov}&[
        (\mathbf{z}_{t_1}-\frac{\sum_{k=1}^\top   c_{k}}{T}\mathbf{1})^\top  B\mathbf{z}_{t_1},
        (\mathbf{z}_{t_2}-\frac{\sum_{k=1}^\top   c_{k}}{T}\mathbf{1})^\top  B\mathbf{z}_{t_2}].
        \end{aligned}
\end{equation}

For simplicity, we use $\mathbf{u},\mathbf{v}$ to denote $\mathbf{z}_{t_1},\mathbf{z}_{t_2}$, respectively, and use $w = \sum_{t=1}^T   c_t/T$. We continue to expand the covariance term, and get:
\begin{equation}
    \sum_{i,j,k,l} B_{i j} B_{k l}
    \left(
        \mathbb{E}\left[\left(u_i-w\right) u_j\left(v_k-w\right) v_l\right]-\mathbb{E}\left[\left(u_i-w\right) u_j\right] \mathbb{E}\left[\left(v_k-w\right) v_l\right]
    \right),
\end{equation}
Correspondingly, we need to discuss the order of the this term in five cases, as well as that in the Appendix~\ref{proof_of_theo1}. Similarly, for the first four cases (except for the case where $i,j,k,l$ are not equal to each other), we allow for:
\begin{equation}
    \mathbb{E}\left[\left(u_i-w\right) u_j\left(v_k-w\right) v_l\right]
    -
    \mathbb{E}\left[\left(u_i-w\right) u_j\right] 
    \mathbb{E}\left[\left(v_k-w\right) v_l\right]
    = \Theta(1),
\end{equation}

Therefore, we only need to prove:
\begin{equation}
    \mathbb{E}\left[\left(u_i-w\right) u_j\left(v_k-w\right) v_l\right]
    -
    \mathbb{E}\left[\left(u_i-w\right) u_j\right] 
    \mathbb{E}\left[\left(v_k-w\right) v_l\right] 
    = O\left(\frac{1}{n}\right),
\end{equation}
when $i,j,k,l$ are not equal to each other. 

Since $\mathbf{u}$ is the treatment vector that is assigned before $\mathbf{v}$, we should first condition on $u_i,u_j$ . 
\begin{equation}
    \begin{aligned}
        \mathbb{E}\left[(u_i-w) u_j (v_k-w) v_l\right] 
        &= 
        \left(c_{t_1}^2 + O\left(\frac{1}{n}\right)\right)(1-w) \mathbb{E}[(v_k-w) v_l \mid u_i=u_j=1] 
        \\
        &\phantom{=}\ \ - \left(c_{t_1}(1-c_{t_1}) + O\left(\frac{1}{n}\right)\right)w \mathbb{E}[(v_k-w) v_l \mid u_j=1] 
        \\
        &=(1-w)c_{t_1}^2 (c_{t_2}-w)c_{t_2} - wc_{t_1}(1-c_{t_1})(c_{t_2}-w)c_{t_2} + O\left(\frac{1}{n}\right) 
        \\
        &=
        (c_{t_1} -w)c_{t_1}(c_{t_2} -w)c_{t_2} + O\left(\frac{1}{n}\right).
        \end{aligned}
\end{equation}

On the other hand, it is easy to check that 
\begin{equation}
    \mathbb{E}\left[\left(u_i-w\right) u_j\right] 
    \mathbb{E}\left[\left(v_k-w\right) v_l\right] = (c_{t_1} -w)c_{t_1}(c_{t_2} -w)c_{t_2} + O\left(\frac{1}{n}\right),        
\end{equation}

Thus, we finish the proof. \qed

\begin{remark}
	
In fact, the covariance term in other four cases are likely of constant order. For example, we check the case where only $j=l$ holds:
\begin{equation}
    \begin{aligned}
        \mathbb{E}
        \left[
        (u_i-w)u_j(v_k-w)v_l
        \right]
        &=
        \mathbb{E}
        \left[
        (u_i-w)u_j(v_k-w)v_j
        \right]
        \\
        &= 
        c_{t_1}\mathbb{E}
        \left[
        (u_i-w)(v_k-w)v_j
        \mid u_j=1 \right]
        \\
        &=
        c_{t_1}\left(c_{t_1}-w + O\left(\frac{1}{n}\right)\right) \left(c_{t_2}-w + O\left(\frac{1}{n}\right)\right)
        \\
        &=
        c_{t_1}(c_{t_1}-w)(c_{t_2}-w) + O\left(\frac{1}{n}\right).
    \end{aligned}
\end{equation}

Therefore, in this case, the value of covariance term 
\begin{equation}
    \begin{aligned}
    &\phantom{=}\ \ \mathbb{E}\left[\left(u_i-w\right) u_j\left(v_k-w\right) v_l\right]
    -
    \mathbb{E}\left[\left(u_i-w\right) u_j\right] 
    \mathbb{E}\left[\left(v_k-w\right) v_l\right] 
    \\
    &=
    c_{t_1}(c_{t_1}-w)(c_{t_2}-w) - c_{t_1}c_{t_2}(c_{t_1}-w)(c_{t_2}-w) + O\left(\frac{1}{n}\right)
    \\ 
    &= \Theta(1),
    \end{aligned}     
\end{equation}
if $c_{t_1}>0$ and $c_{t_2}<1$. Therefore, the third part of Assumption~\ref{assumption1} is still necessary here. 
\end{remark}

\newpage
% Removed empty Simulation Details placeholder.

\section*{NeurIPS Paper Checklist}

%%% END INSTRUCTIONS %%%

\begin{enumerate}

\item {\bf Claims}
    \item[] Question: Do the main claims made in the abstract and introduction accurately reflect the paper's contributions and scope?
    \item[] Answer: \answerYes{} % Replace by \answerYes{}, \answerNo{}, or \answerNA{}.
    \item[] Justification: The abstract and introduction state the main contributions, assumptions, and scope, and these are supported by the theory and experiments in Sections 1--5 and the appendix.

\item {\bf Limitations}
    \item[] Question: Does the paper discuss the limitations of the work performed by the authors?
    \item[] Answer: \answerYes{} % Replace by \answerYes{}, \answerNo{}, or \answerNA{}.
    \item[] Justification: The paper discusses key limitations and scope restrictions, including the focus on static GATE, suppression of temporal interference, and the simplifying assumptions behind the exact theory; see Sections 3--6 and Appendix A.

\item {\bf Theory assumptions and proofs}
    \item[] Question: For each theoretical result, does the paper provide the full set of assumptions and a complete (and correct) proof?
    \item[] Answer: \answerYes{} % Replace by \answerYes{}, \answerNo{}, or \answerNA{}.
    \item[] Justification: The theoretical results are stated with explicit assumptions and cross-referenced proofs, with detailed derivations deferred to the appendix and supplemental material.

    \item {\bf Experimental result reproducibility}
    \item[] Question: Does the paper fully disclose all the information needed to reproduce the main experimental results of the paper to the extent that it affects the main claims and/or conclusions of the paper (regardless of whether the code and data are provided or not)?
    \item[] Answer: \answerYes{} % Replace by \answerYes{}, \answerNo{}, or \answerNA{}.
    \item[] Justification: The paper specifies the experimental design, datasets, models, and evaluation protocol, and we will further ensure reproducibility by providing runnable code in the supplementary material.

\item {\bf Open access to data and code}
    \item[] Question: Does the paper provide open access to the data and code, with sufficient instructions to faithfully reproduce the main experimental results, as described in supplemental material?
    \item[] Answer: \answerYes{} % Replace by \answerYes{}, \answerNo{}, or \answerNA{}.
    \item[] Justification: We will provide anonymized code and instructions in the supplementary material so that the main experimental results can be faithfully reproduced.

\item {\bf Experimental setting/details}
    \item[] Question: Does the paper specify all the training and test details (e.g., data splits, hyperparameters, how they were chosen, type of optimizer) necessary to understand the results?
    \item[] Answer: \answerYes{} % Replace by \answerYes{}, \answerNo{}, or \answerNA{}.
    \item[] Justification: The paper specifies the network data, interference models, merging schemes, randomization schemes, and main model choices in Section 5 and Appendix B, and the supplementary code will include full training details.

\item {\bf Experiment statistical significance}
    \item[] Question: Does the paper report error bars suitably and correctly defined or other appropriate information about the statistical significance of the experiments?
    \item[] Answer: \answerYes{} % Replace by \answerYes{}, \answerNo{}, or \answerNA{}.
    \item[] Justification: The experiments report variability information through standard deviations, and the paper explicitly states that the bias and standard deviation are computed from Monte Carlo repetitions; see Section 5.2 and the result tables.

\item {\bf Experiments compute resources}
    \item[] Question: For each experiment, does the paper provide sufficient information on the computer resources (type of compute workers, memory, time of execution) needed to reproduce the experiments?
    \item[] Answer: \answerNo{} % Replace by \answerYes{}, \answerNo{}, or \answerNA{}.
    \item[] Justification: Computational cost is not a central concern of this paper, whose main focus is the statistical properties of regression-based estimators under network interference. The experiments use lightweight models and standard simulation pipelines, and we will provide implementation details in the supplementary material.
    
\item {\bf Code of ethics}
    \item[] Question: Does the research conducted in the paper conform, in every respect, with the NeurIPS Code of Ethics \url{https://neurips.cc/public/EthicsGuidelines}?
    \item[] Answer: \answerYes{} % Replace by \answerYes{}, \answerNo{}, or \answerNA{}.
    \item[] Justification: To the best of our knowledge, the research complies with the NeurIPS Code of Ethics and does not involve unethical data collection or human-subject procedures.

\item {\bf Broader impacts}
    \item[] Question: Does the paper discuss both potential positive societal impacts and negative societal impacts of the work performed?
    \item[] Answer: \answerNo{} % Replace by \answerYes{}, \answerNo{}, or \answerNA{}.
    \item[] Justification: The current submission does not include a dedicated discussion of broader societal impacts. This is a methodological paper on experimental design and estimation.
    
\item {\bf Safeguards}
    \item[] Question: Does the paper describe safeguards that have been put in place for responsible release of data or models that have a high risk for misuse (e.g., pre-trained language models, image generators, or scraped datasets)?
    \item[] Answer: \answerNA{} % Replace by \answerYes{}, \answerNo{}, or \answerNA{}.
    \item[] Justification: The paper does not release high-risk models or scraped datasets, so special safeguards for responsible release are not applicable here.

\item {\bf Licenses for existing assets}
    \item[] Question: Are the creators or original owners of assets (e.g., code, data, models), used in the paper, properly credited and are the license and terms of use explicitly mentioned and properly respected?
    \item[] Answer: \answerNo{} % Replace by \answerYes{}, \answerNo{}, or \answerNA{}.
    \item[] Justification: Existing assets such as the FB-Stanford3 network and software packages are credited by citation, but the manuscript does not yet explicitly document licenses and terms of use. We will add this information in the supplementary material.

\item {\bf New assets}
    \item[] Question: Are new assets introduced in the paper well documented and is the documentation provided alongside the assets?
    \item[] Answer: \answerYes{} % Replace by \answerYes{}, \answerNo{}, or \answerNA{}.
    \item[] Justification: We will release code for the proposed method and experiments, together with documentation and instructions in the supplementary material.

\item {\bf Crowdsourcing and research with human subjects}
    \item[] Question: For crowdsourcing experiments and research with human subjects, does the paper include the full text of instructions given to participants and screenshots, if applicable, as well as details about compensation (if any)? 
    \item[] Answer: \answerNA{} % Replace by \answerYes{}, \answerNo{}, or \answerNA{}.
    \item[] Justification: The paper does not involve crowdsourcing or experiments with human subjects.

\item {\bf Institutional review board (IRB) approvals or equivalent for research with human subjects}
    \item[] Question: Does the paper describe potential risks incurred by study participants, whether such risks were disclosed to the subjects, and whether Institutional Review Board (IRB) approvals (or an equivalent approval/review based on the requirements of your country or institution) were obtained?
    \item[] Answer: \answerNA{} % Replace by \answerYes{}, \answerNo{}, or \answerNA{}.
    \item[] Justification: The paper does not involve human subjects, so IRB approval or an equivalent review is not applicable.

\item {\bf Declaration of LLM usage}
    \item[] Question: Does the paper describe the usage of LLMs if it is an important, original, or non-standard component of the core methods in this research? Note that if the LLM is used only for writing, editing, or formatting purposes and does \emph{not} impact the core methodology, scientific rigor, or originality of the research, declaration is not required.
    %this research? 
    \item[] Answer: \answerNA{} % Replace by \answerYes{}, \answerNo{}, or \answerNA{}.
    \item[] Justification: LLMs are not an important or non-standard component of the core methodology in this research; any LLM use was limited to writing assistance only, if any.

\end{enumerate}

\end{document}